\documentclass[twocolumn,showpacs,pre]{revtex4}
\usepackage{amsmath}
\usepackage{graphicx}
\usepackage{bm}
\usepackage{color}
\usepackage{amsfonts}
\usepackage{dcolumn}

\begin{document}
\title{Analysis of a growing dynamic length scale in a glass-forming binary hard-sphere mixture}

\author{Elijah Flenner, Min Zhang, and Grzegorz Szamel}
\affiliation{Department of Chemistry, Colorado State University, Fort Collins, CO 80523}
\date{\today}

\begin{abstract}
We examine a length scale that characterizes the spatial extent of heterogeneous dynamics in a glass-forming binary 
hard-sphere mixture up to the mode-coupling volume fraction $\phi_c$. 
First, we characterize the system's dynamics. Then, we utilize a new method
[Phys.\ Rev. Lett.\ \textbf{105}, 217801 (2010)] to extract and analyze the ensemble independent dynamic 
susceptibility $\chi_4(t)$ and the dynamic correlation length $\xi(t)$ for a range of times between the $\beta$ 
and $\alpha$ relaxation times. We find that in this time range 
the dynamic correlation length follows a volume fraction independent curve
$\xi(t) \sim \ln(t)$. For longer times, $\xi(t)$ departs from this curve and remains constant
up to the largest time at which we can determine the length accurately. 
In addition to the previously established correlation $\tau_\alpha \sim \exp[\xi(\tau_\alpha)]$ 
between the $\alpha$ relaxation time, $\tau_\alpha$, and the 
dynamic correlation length at this time, $\xi(\tau_\alpha)$,
we also find a similar correlation for the diffusion coefficient 
$D \sim \exp[\xi(\tau_\alpha)^\theta]$ with $\theta \approx 0.6$. 
We discuss the relevance of these findings for different theories of the glass transition. 
 
\end{abstract}


\maketitle

\section{Introduction}\label{introduction}
It is becoming increasingly apparent that growing length scales can be associated with the dramatic slowing
down of the dynamics in glass-forming systems. One such length scale 
characterizes the spatial extent of the so-called dynamic heterogeneity.
It has been found that upon approaching the glass transition the particles' motion becomes increasingly heterogeneous 
and the particles can be divided into "slow" and "fast" sub-sets
\cite{Sillescu1999,Ediger2000,Glotzer2000,Richert2002,Andersen2005}. 
These sub-sets can be seen as distinct peaks in the probability of the logarithm of single particle displacements 
$P[\log_{10}(\delta r);t]$ \cite{Flenner2005sim,Reichman2005,Saltzman2008}.
Importantly, the
slow and fast particles are not uniformly distributed in space, but form clusters whose size increases
as the dynamics slows.  
The average spatial extent of the clusters of slow particles can be defined as a dynamic correlation length.
This dynamic correlation length and other closely related lengths have been studied in simulations 
\cite{Donati1999,Lacevic2003,Whitelam2004,Berthier2004,Stein2008,Flenner2009,Karmakar2009,Karmakar2010}, experiments
\cite{Berthier2005,Dalle-Ferrier2007,CrausteThibierge2010}, 
and discussed theoretically
\cite{Biroli2004,Biroli2006,Berthier2007p2,Berthier2007p3,Szamel2008,Szamel2010}. 

One convenient way to characterize the spatial extent of the clusters is to identify the slow particles and then 
determine their spatial correlations. In simulational investigations this is typically done by analyzing the 
so-called four-point dynamic structure factors $S_4(q;t)$. The four-point structure factor quantifies spatial 
correlations between the slow particles. The label ``four-point'' refers to the fact that $S_4(q;t)$ is a 
correlation function of two two-point functions that 
are used to characterize particles' dynamics and to define slow particles. Examples of these two point functions
are the microscopic intermediate scattering functions and overlap functions, but other functions have also
been used in the literature. Of particular interest are two quantities that can be expressed
in terms of $S_4(q;t)$: the dynamic susceptibility, 
$\chi_4(t) = \lim_{q\rightarrow 0} S_4(q;t)$, which is a measure of the overall strength of the dynamic heterogeneity, 
and the dynamic correlation length, $\xi(t)$, which characterizes its spatial extent.
The relationship between $\chi_4(t)$ and $\xi(t)$ provides insight into the fractal dimension of the slow particles clusters.
 
In spite of a relatively straightforward definition of $\chi_4(t)$, its 
direct simulational evaluation suffers from a technical difficulty. 
The difficulty originates from the fact that in a typical simulational ensemble some global fluctuations are
suppressed. For example, in most simulations of glass-forming liquids
the number of particles and the volume is kept constant, thus the density of the system is constant and 
global density fluctuations do not contribute to the direct simulational calculation of $\chi_4(t)$. 
Berthier \textit{et al.}\ \cite{Berthier2005} proposed that these suppressed fluctuations
can be calculated utilizing a procedure derived by Lebowitz \textit{et al.} \cite{Lebowitz1967}. 
This procedure results in a two-part expression for the dynamic susceptibility, 
$\chi_4(t) = \chi_4(t)|_{\mathbf{x}} + \mathcal{X}(t)$ where $\chi_4(t)|_{\mathbf{x}}$ is the susceptibility in an ensemble 
with $\mathbf{x}$ kept fixed (which can readily be obtained from simulations) and $\mathcal{X}(t)$ is a correction term.

Berthier \textit{et al.}\ \cite{Berthier2005} furthermore noted that while $\chi_4(t)|_{\mathbf{x}}$ cannot be easily
determined experimentally, the correction term $\mathcal{X}(t)$ can. This fact, together with the positive definite character
of the former term, $\chi_4(t)|_{\mathbf{x}} > 0$, provides an experimental 
lower bound for $\chi_4(t)$. However, the relative size of $\chi_4(t)$ and the correction term remained an open question.
In addition, even if the experimental lower bound was a good estimate of $\chi_4(t)$, there was no reliable correlation
between $\chi_4(t)$ and the dynamic correlation length $\xi(t)$.

The relative size of the two terms contributing to $\chi_4(t)$ was investigated by Berthier \textit{et al.} 
\cite{Berthier2007p2,Berthier2007p3} and by Brambilla \textit{et al.} \cite{Brambilla2009}. The main conclusion
was that as the dynamics slows, the correction term becomes an increasingly
better approximation for the ensemble independent $\chi_4(t)$. The assessment and extension
of this result is one of the subjects of the present paper.

There were several earlier simulational investigations \cite{Lacevic2003,Whitelam2004,Berthier2004,Stein2008,Flenner2009,Karmakar2009}
of the correlation between the dynamic susceptibility $\chi_4(t)$ and the 
correlation length $\xi(t)$ and between the average dynamics (as characterized by, \textit{e.g.} the $\alpha$
relaxation time, $\tau_\alpha$) and $\xi(t)$, but their results, by and large, disagreed \cite{Karmakar2010c}. 
Recently, it has been realized that in order to get reliable results for $\chi_4(t)$ and $\xi(t)$ 
one has to simulate systems considerably larger than was customary 
\cite{Karmakar2010,Karmakar2010c}.
In an earlier short note \cite{Flenner2010prl}, we described an application of 
the method of Lebowitz \textit{et al.}\ \cite{Lebowitz1967} to facilitate
the determination of $\xi(\tau_\alpha)$ using large scale, $80\, 000$ 
particles, simulations. We found that $\xi(\tau_\alpha) \sim \ln(\tau_\alpha)$ over the full range of 
densities studied. This slower, logarithmic growth of $\xi(\tau_\alpha)$ with $\tau_\alpha$ is 
more consistent with experimental findings than the power law growth found in many previous simulations.

In the present paper we give details omitted in Ref.~\cite{Flenner2010prl} due to length restrictions.
In addition, we analyze the time dependence of both the dynamic susceptibility and correlation length,
and investigate additional correlations between the length and the average dynamics. 

We describe the
system and simulation method in Sec.~\ref{simulation} and we briefly characterize the system's dynamics 
in Sec.~\ref{dynamics}. Since there is no accepted theory of the glass transition, examination 
of the system's dynamics can get bogged down with an extensive number of different fits and
fit parameters. Throughout much of the paper we refer to two regimes: 
a mode-coupling like regime where power laws describe the data well, and 
a different dynamic regime where the mode-coupling like power laws are not applicable.
We discuss this characterization of the data and describe the relevant fits in Appendix \ref{fits}.
After describing the system's dynamics, in Sec.~\ref{length} we investigate the dynamic susceptibility $\chi_4(t)$ and the 
dynamic correlation length $\xi(t)$. The technical details of the calculation of $\chi_4(t)$ and $\xi(t)$ 
are given in Appendix \ref{xicalc}. In Sec.~\ref{lengthdynamics} we explore the connections
between the dynamic correlation length and the average dynamics. We finish with a discussion in Sec.~\ref{conclusions}.
 
\section{Simulation Details}\label{simulation}
We simulated a system introduced by Brambilla \textit{et al.}\ \cite{Brambilla2009}: 
a 50:50 binary hard-sphere mixture where the diameter $d_2$ of the larger sphere is
$1.4$ times larger than the diameter $d_1$ of the smaller sphere. The size difference is chosen to inhibit
crystallization. We studied systems with $N = (N_1 + N_2) = 80\, 000$ particles 
at volume fractions  $\phi= \pi (N_1 d_1^3 + N_2 d_2^3)/6 V$ equal to 0.4, 0.45, 0.5, 0.52, 0.55, 0.56, 0.57, 
0.58, and 0.59 and systems with $10\, 000$ particles at volume fractions $\phi$ equal to 0.54, 0.575, 0.58, 
0.585, and 0.59. Additional simulations were performed at slightly different volume fractions
and concentrations to obtain the derivatives needed in this work. To determine the derivatives 
with respect to $\phi$, we performed simulations at $\phi \pm \delta\phi$ where $\delta \phi = 0.001$ 
for $\phi \le 0.58$ and $\delta \phi = 0.0005$ for $\phi \ge 0.585$. To determine the derivatives with
respect to concentration $c = N_1/N$, we performed simulations at $c = 0.5 \pm 0.05$ for
$\phi \le 0.58$. The concentration derivatives had very little $\phi$ dependence over the range we examined. 
We found that they were not necessary to obtain accurate correlation lengths and susceptibilities 
for $\phi \ge 0.56$. Thus we did not determine the concentration derivatives for $\phi \ge 0.585$.

We performed Monte Carlo simulations with the local trial displacements of particles randomly 
chosen from a cube of length $0.1 d_1$. It has been shown that Monte Carlo dynamics reproduces well the long time 
dynamics of glass forming systems \cite{Berthier2007p4}. Moreover, Brambilla \textit{et al.}\ \cite{Brambilla2009} 
have shown that the present system with this particular Monte Carlo dynamics reproduces well the 
long-time dynamics of their experimental system - a dense poly-disperse hard sphere system 
in which hydrodynamic interactions can be neglected.  

The simulations were run for at least $100 \tau_\alpha$
($\tau_\alpha$ is defined in Section \ref{dynamics}) after the systems stopped aging. 
To check for the presence of aging, we examined two point and four point quantities to see if they significantly
depended on the initial time of the calculation. We found that the dynamic susceptibility $\chi_4(t)|_{\phi,c}$
is very sensitive to aging, thus providing a good test of equilibration. We ran at least four production runs
at each volume fraction, and the results are an average over those runs.  Results are presented in reduced
units where the unit of length is $d_1$ and the unit of time $t$ is one Monte Carlo step (a Monte Carlo step
is one attempted move per particle).  Since the center of mass of the system can drift, all positions are calculated with respect
to the center of mass (for the calculation of the center of mass position masses of all the particle were taken as identical). 

\section{Single Particle Dynamics}\label{dynamics}
In this section we examine the slowing down of the average dynamics. In addition, we show that there is an 
indication of dynamic heterogeneity in two-point functions, and the dynamic heterogeneity
 is increasing with volume fraction. 

We start
by examining the volume fraction dependence of the $\alpha$ relaxation time, $\tau_\alpha$, determined
by a characteristic decay time of an average overlap function. The average overlap function is defined as 
\begin{equation}
F_o(t) =  \frac{1}{N} \left< \sum_{n=1}^N w_n(t)\right>,
\end{equation}
where $w_n(t)$ is a microscopic overlap function,
\begin{equation}\label{mof}
w_n(t) = \Theta[a-|\mathbf{r}_n(t) - \mathbf{r}_n(0)|].
\end{equation}
Here $\Theta(x)$ is Heaviside's step function and $\mathbf{r}_n(t)$ is the position of particle $n$ at
a time $t$. The microscopic overlap function $w_n(t)$ select particles that did not move farther than $a$ from their
original positions during the time $t$. In this work we use $a=0.3$. Correspondingly, the average overlap
function $F_o(t)$ measures the average fraction of particles which did not move farther than $a$ from their
original positions during the time $t$. We will refer to particles which did not move farther than $a$ during 
time $t$ as the slow particles. Thus, 
\begin{equation}\label{nosp}
N_s(t) = \sum_{n=1}^N w_n(t)
\end{equation} 
is the number of slow particles during time $t$, and 
$\left<N_s(t) \right> = N F_o(t)$ is the average number of slow particles 
during time $t$.

$F_o(t)$ encodes similar information as the self intermediate
scattering function $F_s(q;t) = N^{-1} \left< \sum_n e^{-\mathbf{q} \cdot [\mathbf{r}_n(t) - \mathbf{r}_n(0)]} \right>$. Thus, 
it displays similar characteristics. At high densities a pleateau region develops in the time dependence of $F_o(t)$. Moreover, 
an early $\beta$ relaxation regime can be identified as
the decay to the plateau, and then the late $\beta$ regime can be seen as a decay from the plateau. The characteristic time
of the final decay from the plateau 
is referred to as the $\alpha$ relaxation time $\tau_\alpha$. We define $\tau_\alpha$ adopting the formula used before
for the self-intermediate scattering function, $F_o(\tau_\alpha) = e^{-1}$. Consequently, according to this definition 
the average fraction of slow particles during time $\tau_\alpha$ is about 37\%.

\begin{figure}
\includegraphics[width=3.2in]{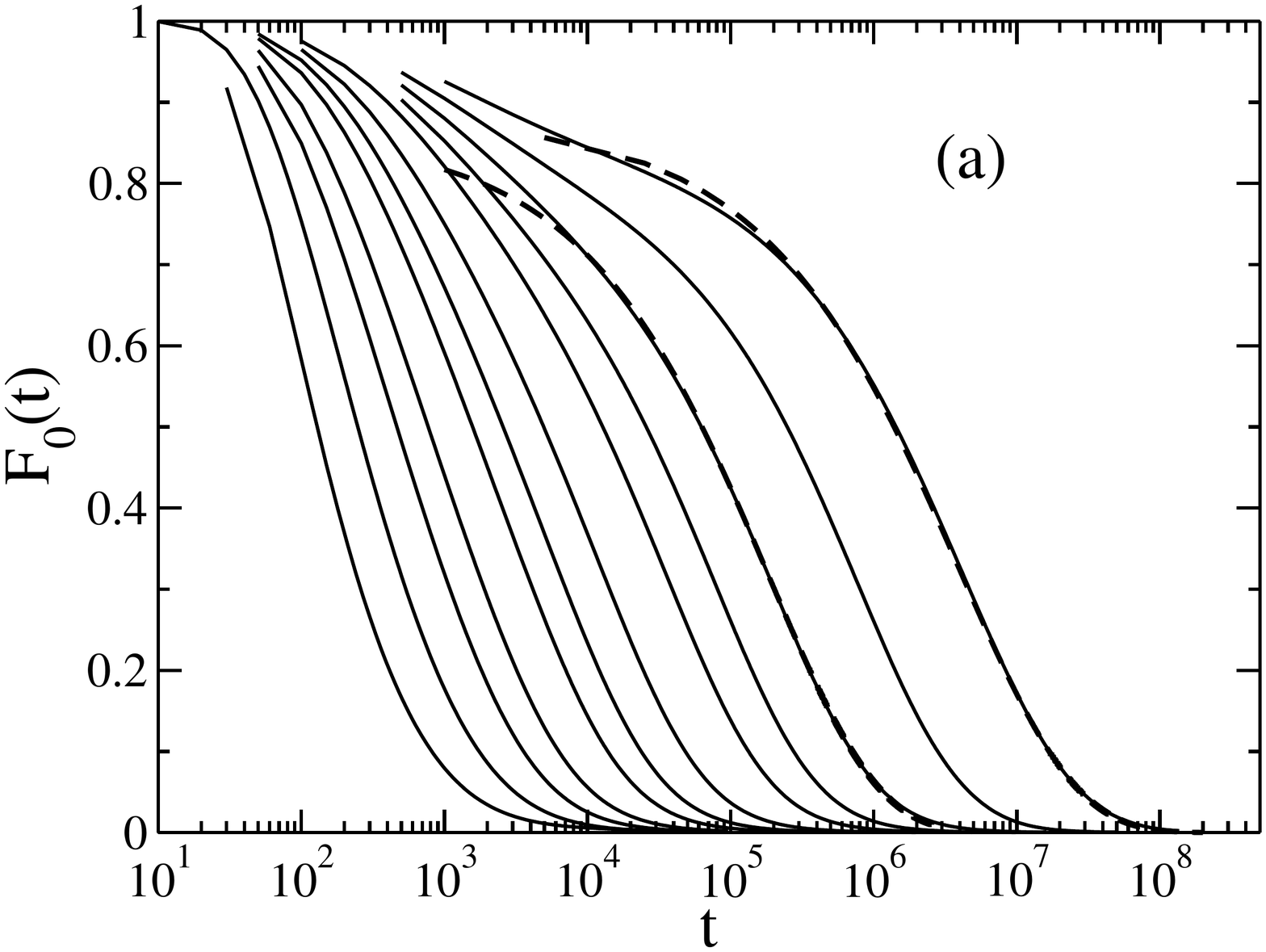}
\includegraphics[width=3.2in]{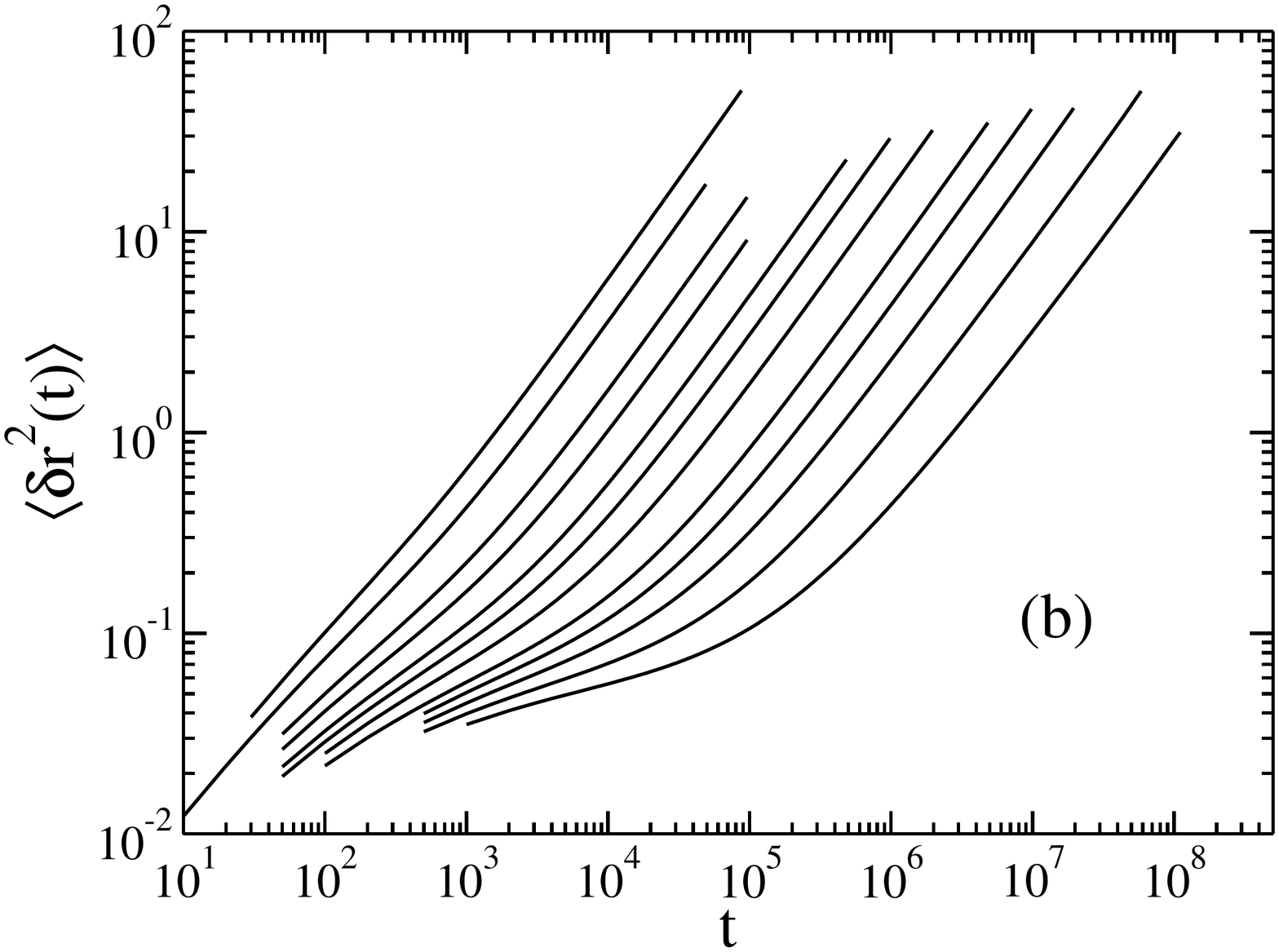}
\caption{\label{fig:Fo}(a): The average overlap function $F_o(t)$ for $\phi = 0.4$, 0.45, 0.5, 0.52, 0.54, 0.55, 
0.56, 0.57, 0.575, 0.58, 0.585, and 0.59 listed from left to right. 
The dashed lines are stretched exponential, $\exp[ -(t/\tau)^{\beta} ]$, 
fits at $\phi = 0.58$ ($\beta = 0.56$) and $\phi = 0.59$ ($\beta = 0.54$). (b): The mean square displacement 
$\left< \delta r^2(t) \right>$ for $\phi = 0.4$, 0.45, 0.5, 0.52, 0.54, 0.55, 
0.56, 0.57, 0.575, 0.58, 0.585, and 0.59 listed from left to right. }
\end{figure}

Shown in Fig.~\ref{fig:Fo}(a) is $F_o(t)$ for $\phi = 0.4$, 0.45, 0.5, 0.52, 0.54, 0.55, 0.56, 0.57, 0.575, 0.58, 0.585, 
and 0.59 listed from left to right. For small volume fractions the decay is nearly exponential. At higher volume
fractions the long time decay follows a stretched exponential
form $\exp[ -(t/\tau)^{\beta} ]$ with a  weakly $\phi$ dependent $\beta \approx 0.55$. 
The stretched exponential relaxation is usually interpreted as an indication of dynamic heterogeneity. 

The other common way to examine the average dynamics is to investigate the mean square displacement
displacement 
\begin{equation}
\left< \delta r^2(t) \right> = N^{-1} \left< \sum_n [ \mathbf{r}_n(t) - \mathbf{r}_n(0)]^2 \right>,
\end{equation}
which is shown in Fig.~\ref{fig:Fo}(b). Again, for large $\phi$ a plateau forms at intermediate times,
then at long times $\left< \delta r^2(t) \right> = 6 D t$ where
$D$ is the self diffusion coefficient.  Both the previously mentioned 
plateau in the average overlap function $F_o(t)$ and the plateau in 
the mean square displacement $\left< \delta r^2(t) \right>$ are associated with the so-called cage effect
where particles are temporarily trapped by cages of neighboring particles. 
We use the long time limiting behavior of $\left< \delta r^2(t) \right>$ to obtain the self diffusion coefficient $D$.  
We define the $\beta$ relaxation time $\tau_\beta$ as the inflection point of $\ln[\left< \delta r^2(t) \right>]$ versus $\ln(t)$
(we found that it is easier to determine $\tau_\beta$ from the $\ln[\left< \delta r^2(t) \right>]$ inflection point rather
than from the $F_o(t)$ inflection point). 
This inflection point could only be determined for $\phi \ge 0.5$. 

\begin{figure}
\includegraphics[width=3.2in]{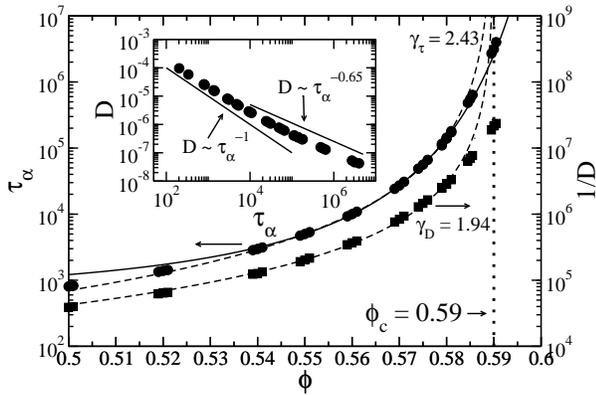}
\caption{\label{tauD}The relaxation time $\tau_\alpha$ and the inverse diffusion coefficient $1/D$. The dashed lines are 
power law fits $a (\phi_{c} - \phi)^{-\gamma}$ and the solid line is a fit suggested by Berthier and Witten \cite{Berthier2009},
$\tau_\infty \exp[ B/(\phi_0 - \phi)^2 ]$. Inset: Stokes-Einstein violation: for small volume fractions $D \sim \tau_\alpha^{-1}$,
whereas for higher volume fraction $D \sim \tau_\alpha^{-0.65}$. }
\end{figure}

Shown in Fig.~\ref{tauD} is the volume fraction dependence
of the relaxation time, $\tau_\alpha$, and the inverse of the self-diffusion coefficient, $1/D$. 
As in many glass forming systems, there is an 
range of volume fractions in which power laws provide good fits to the simulation data. Since
the mode-coupling theory predicts power law divergences of both $\tau_\alpha$ and $1/D$ \cite{Gotze1991}, 
the volume fraction where the power law fits diverge is referred to
as the mode-coupling volume fraction $\phi_{c}$. 
However, there is no true divergence at $\phi_{c}$, the mode-coupling transition is
said to be avoided, and there is emergence of new behavior beyond $\phi_{c}$. To quantitatively identify a 
mode-coupling like region of the dynamics we fit $\tau_\alpha$ and $1/D$ to power laws 
$a (\phi_{c} - \phi)^{-\gamma_{\{\tau,D\}}}$ where $\gamma_\tau$ and $\gamma_D$ denote
the power law exponents for $\tau_\alpha$ and $1/D$, respectively (see Appendix \ref{fits} for
a detailed description of the fits). We found that power laws 
describe our data well for $0.55 \le \phi \le 0.58$ with $\phi_{c} = 0.59$. These
fits are shown as dashed lines in Fig.~\ref{tauD}. 
We also find that our results for the relaxation time 
are consistent with a fit suggested by Berthier and Witten \cite{Berthier2009} and later used by 
Brambilla \textit{et al.}\ \cite{Brambilla2009}, 
$\tau_\alpha = \tau_\infty \exp[ B/(\phi_0 - \phi)^2 ]$. This  
fit is shown as the solid line in Fig.~\ref{tauD}  
(again, see Appendix \ref{fits} for a detailed description of this fit)

In the inset in Fig.~\ref{tauD} we investigate the relation between two quantities discussed above, the $\alpha$ relaxation
time and the self-diffusion coefficient. For small $\phi$ we find that $D \sim \tau_\alpha^{-1}$ and thus 
the Stokes-Einstein relation is obeyed. 
With increasing $\phi$ there appears to be a crossover to a weaker dependence of the self-diffusion coefficient
on the $\alpha$ relaxation time. Thus, the Stokes-Einstein relation is violated. The breakdown of this relations is 
considered to be one of the hallmarks of dynamic heterogeneity.

Quantitatively, we find that for large volume fractions $D \sim \tau_\alpha^{-\sigma}$ where 
$\sigma \approx 0.65$. We should note that for 
even larger $\phi$ it may be found that $\sigma < 0.65$, and our result should be considered an upper bound.
A value of $\sigma = 0.77$ was found in an experimental glass-former \cite{Swallen2003},
and kinetically-constrained lattice-gas models predict a fragility dependent $\sigma$ with values 
between 0.58 and 0.88 \cite{Pan2005}. 
The Random-First-Order theory also predicts a fragility dependent $\sigma$ \cite{Xia2001}.

\begin{figure}
\includegraphics[width=3.2in]{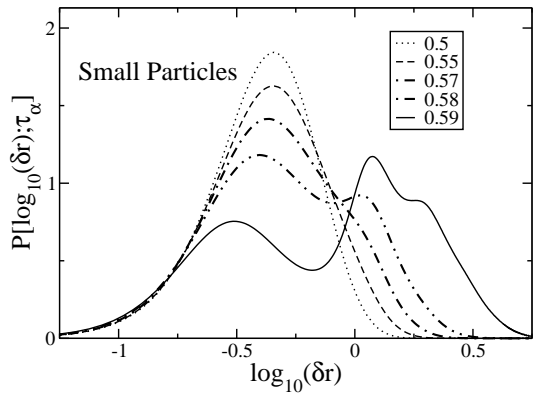}
\caption{\label{plog}The probability of the logarithm of single particle
displacement for the small particles calculated at $\tau_\alpha$ for
several representative volume fractions. For larger volume fractions 
there appears a multi-peak structure of $P[\log_{10}(\delta r);t]$ 
which indicates the existence of sub-populations of slow and fast particles. 
 }
\end{figure}

\begin{figure}
\includegraphics[width=3.2in]{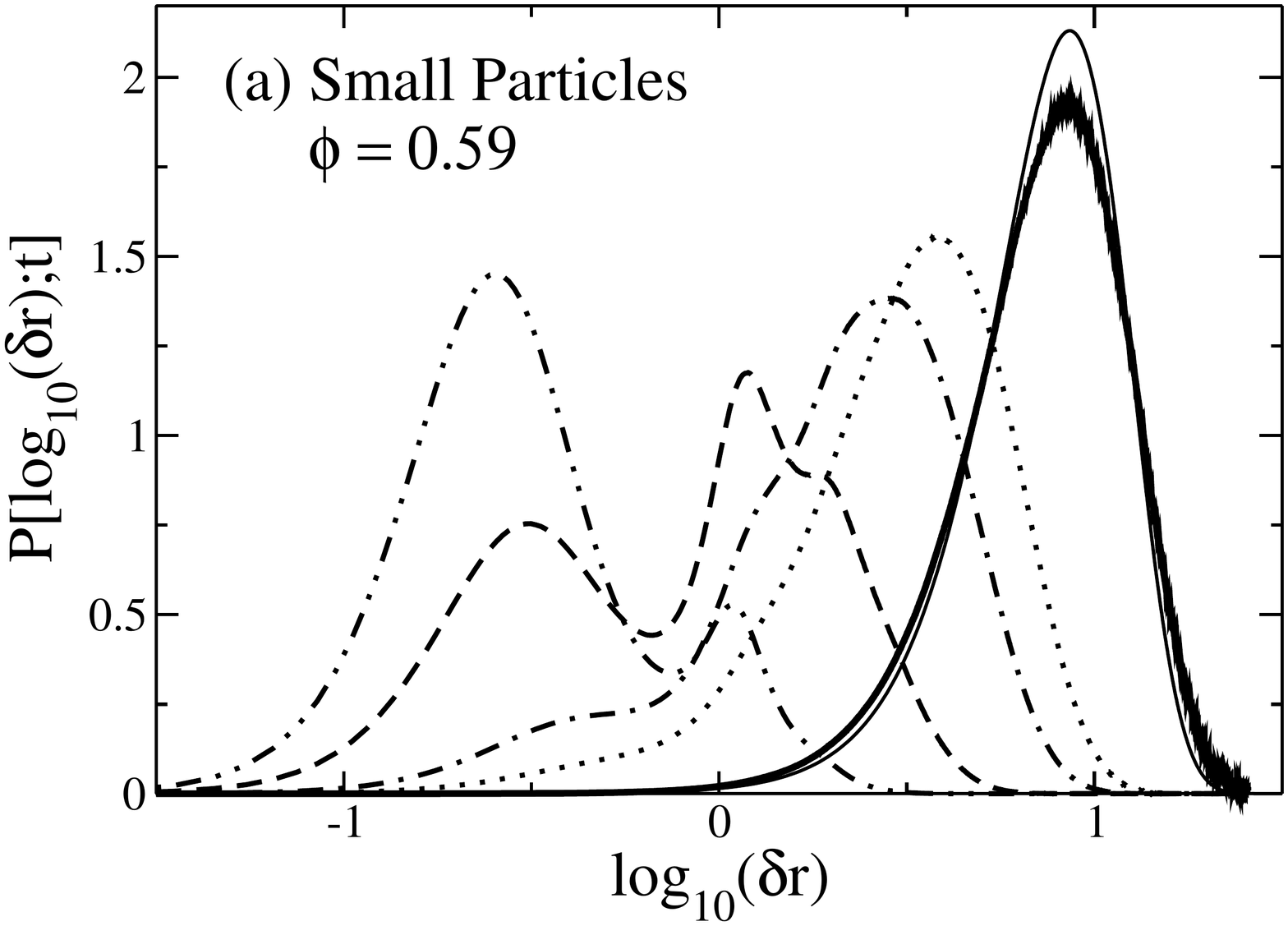}
\includegraphics[width=3.2in]{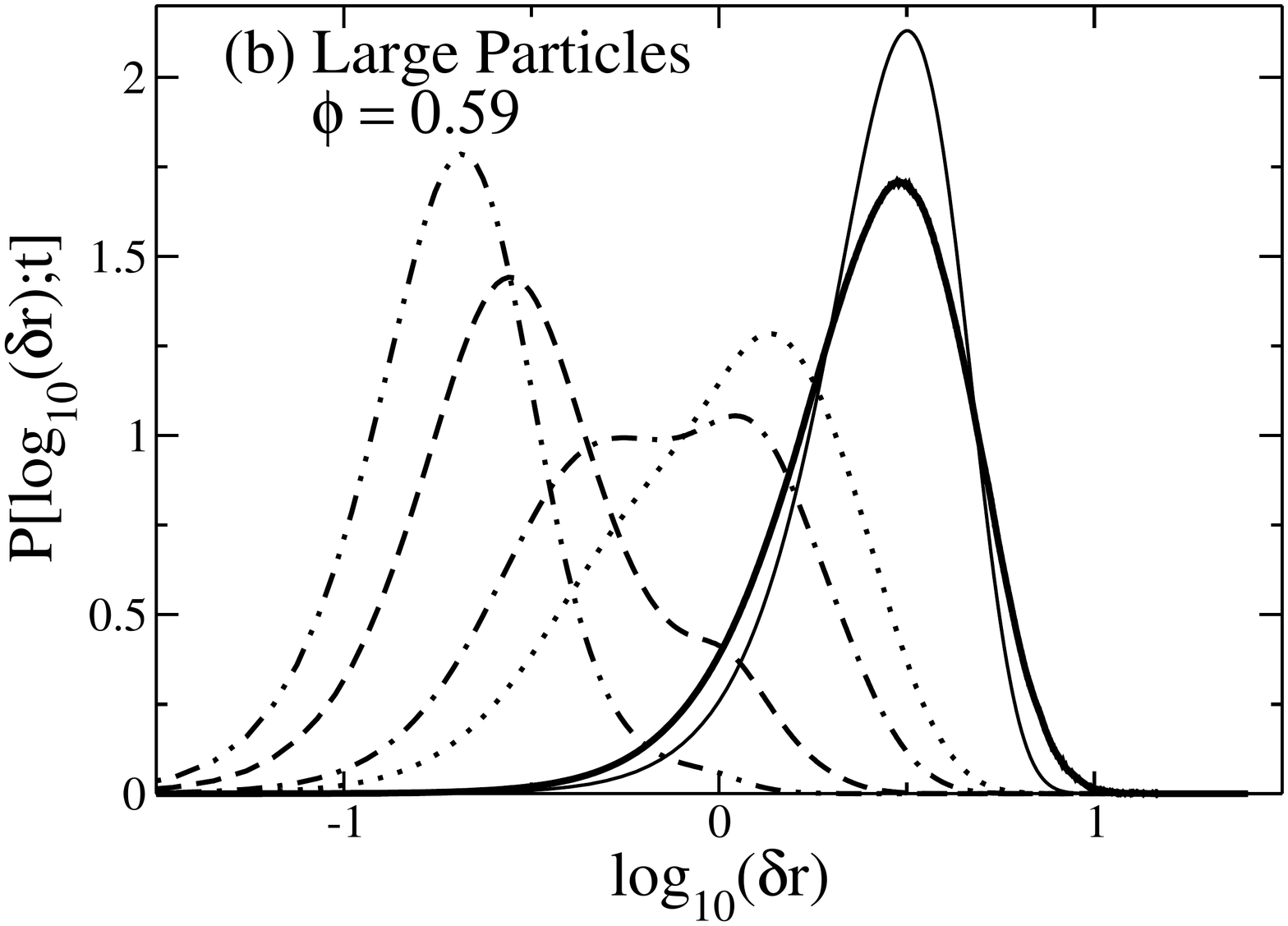}
\caption{\label{plogtime} The probability of the logarithm of single
particle displacements for $\phi = 0.59$ at $0.1 \tau_\alpha$, $\tau_\alpha$, $5 \tau_\alpha$,
$10 \tau_\alpha$, and $50 \tau_\alpha$ listed from left to right. The thin solid line 
is $P[\log_{10}(\delta r);t]$ at $50 \tau_\alpha$ calculated for a Gaussian distribution of displacements 
with the same $\left< \delta r^2 \right>$ as measured in the simulation. (a)
small particles. (b) large particles.}
\end{figure}

To further explore the heterogeneous dynamics we  
examined the probability of the logarithm of single particles displacements, 
$P[\log_{10}(\delta r);t]$ at $\tau_\alpha$. This probability distribution is
related to the self van Hove correlation function, 
$G_s(\delta r;t) = \left< \delta \mathbf{r} - \left[\mathbf{r}_{n}(0) - \mathbf{r}_n(t)\right] \right>$, through the
relationship $P[\log_{10}(\delta r);t] = \ln(10) 4 \pi \delta r^3 G_s(\delta r;t)$. The advantage of
examining $P[\log_{10}(\delta r);t]$ is that for Fickian diffusion (\textit{i.e.} for a Gaussian distribution of
single particle displacements) its shape is 
independent of time and its peak value is constant and approximately equal to 2.13 \cite{Flenner2005sim}. 
Thus, the time-dependence of the shape of $P[\log_{10}(\delta r);t]$ is clear evidence 
of non-Fickian motion. Furthermore, a multi-peak structure of $P[\log_{10}(\delta r);t]$ is indication of the 
presence of distinct sub-populations of particles and, thus, heterogeneous dynamics. 

Shown in Fig.~\ref{plog} is $P[\log_{10}(\delta r);\tau_\alpha]$ calculated for the small particles at 
$\phi = 0.5$, 0.55, 0.57, 0.58, and 0.59. The behavior is similar to what has been observed in 
other simulated glass-formers \cite{Flenner2005sim,Flenner2005mc,Reichman2005,Saltzman2008}
in that multiple peaks emerge close to and at $\phi_{c}$. These peaks correspond to slow and fast particles. 
In Fig.~\ref{plogtime} we show the time dependence of $P[\log_{10}(\delta r);t)]$ for
the small particles (a) and large particles (b) for $\phi = 0.59$. 
Shown are times equal to $0.1 \tau_\alpha$, $\tau_\alpha$, $5 \tau_\alpha$, $10 \tau_\alpha$, and $50 \tau_\alpha$.
The multiple peaks are evident for both types of particles. The peaks are less pronounced and occur at later
times for the large particles. 

For long times we would expect the particles to undergo Fickian diffusion. 
To compare the measured $P[\log_{10}(\delta r);\tau_\alpha]$ with those corresponding to Fickian motion
we show in Fig.~\ref{plog} the probability distributions calculated at $50 \tau_\alpha$ assuming Gaussian
distributions of displacements with the same $\left<\delta r^2(t)\right>$. It is clear that while for the small particles
the difference between the measured distribution and the Fickian one is relatively small, a pronounced difference
is observed for the large particles. Thus, even at the relatively long time, $50 \tau_\alpha$, large particles' motion
is significantly non-Fickian. It should be emphasized that this conclusion cannot be obtained by only investigating the 
time dependence of the mean square displacement which grows approximately linearly with time on this time scale. 
Finally, we recall that in an earlier study we showed that in a binary Lennard-Jones system the time scale
associated with the onset of Fickian diffusion 
increases faster with decreasing temperature than the $\alpha$ relaxation time \cite{Szamel2006}. 
We expect that a corresponding result, \textit{i.e.} that the time scale for the onset of Fickian diffusion 
grows faster with increasing volume fraction than the $\alpha$ relaxation time, holds for the present system.

While one sees clear indications of different sub-populations of slow and fast particles at the higher volume fractions from
the results presented in Figs.~\ref{plog} and \ref{plogtime}, one cannot determine how these slow and fast particles
are distributed in space. In the next section we examine the spatial correlations amongst the
slow  particles. As we mentioned in the opening paragraph of the introduction,
these particles form clusters and a dynamic correlation length can be associated with the average
spatial extent of the clusters.

\section{Dynamic Susceptibility and Correlation Length}\label{length}
\begin{figure}
\includegraphics[width=3.5in]{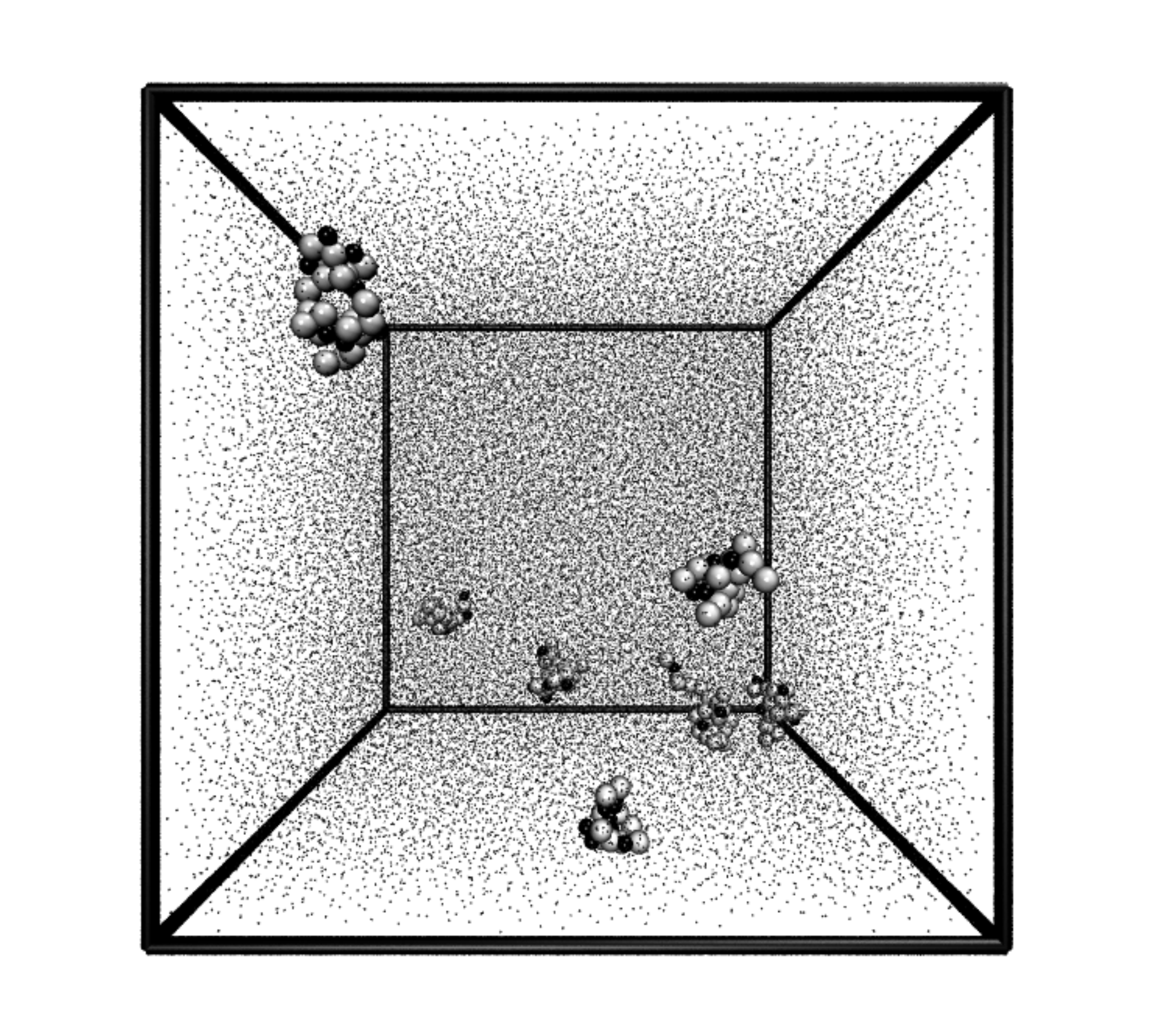}\\
\includegraphics[width=3.5in]{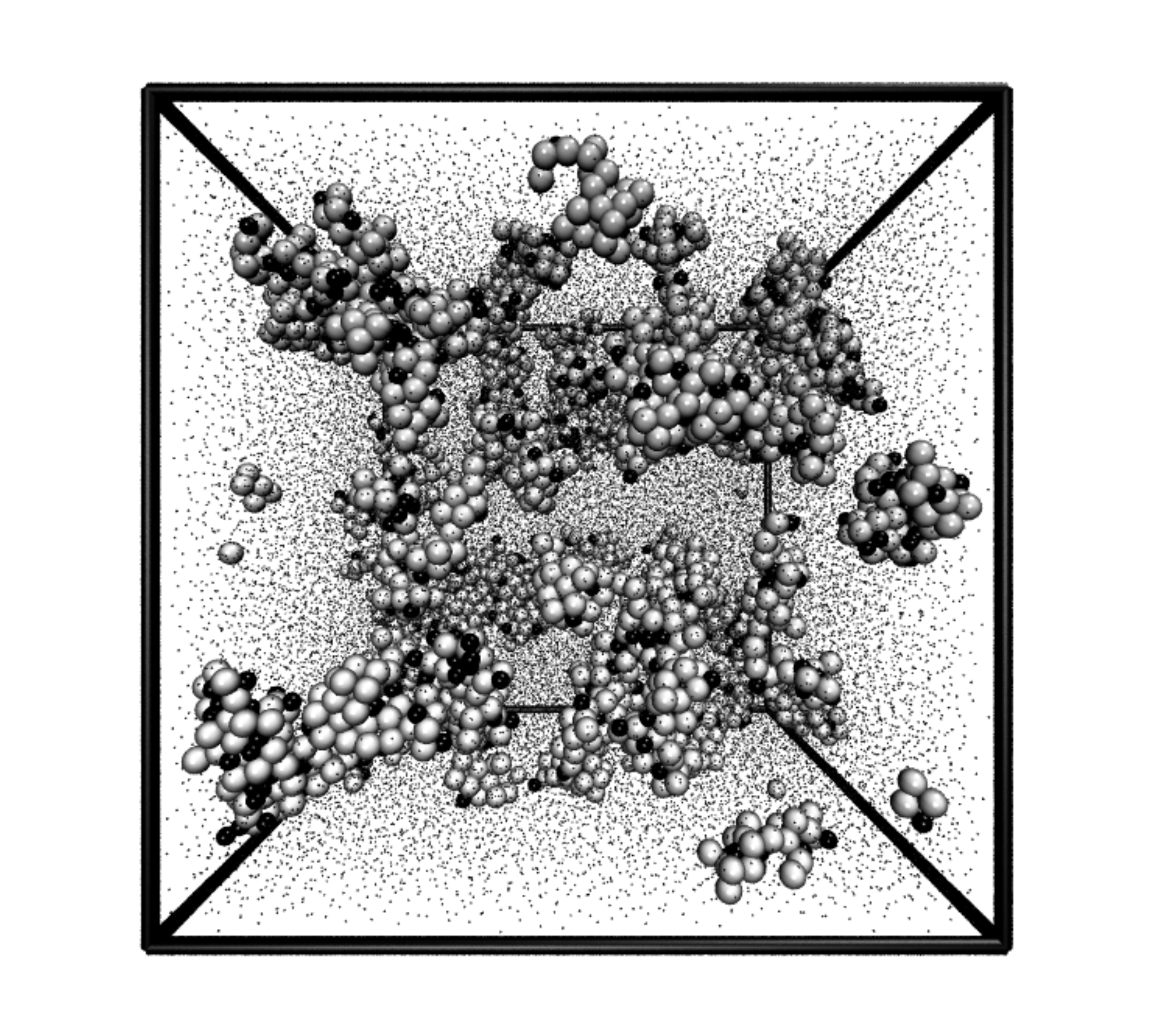}
\caption{\label{slowclusters}Slow particles' 
clusters at $\tau_\alpha$ containing more than 20 particles identified using the algorithm described 
in the text at $\phi = 0.55$ (upper figure) and $\phi = 0.59$ (lower figure).
The white spheres are the large particles, and the black spheres are the small particles. Particles not
belonging to the clusters are shown as black dots.}
\end{figure}

To examine the spatial extent of the heterogeneous dynamics,
we start with a somewhat qualitative approach and look at clusters of slow particles during $\tau_\alpha$ utilizing a somewhat arbitrary definition. 
To visualize these clusters we define the slow particles as those whose displacement
$|\mathbf{r}(t) - \mathbf{r}(0)|$ was less than $a=0.3$ over a time
$t = \tau_\alpha$. We then define two slow particles to be in the same cluster if their initial positions were less than 
$d_{\alpha \beta} + \Delta_{\alpha \beta}$
apart where $d_{\alpha \beta} = (d_\alpha+d_\beta)/2$ and we used $\Delta_{\alpha \beta} = 0.02$. 
Shown in Fig.~\ref{slowclusters} are clusters
of more than 20 slow particles for $\phi=0.55$ and $\phi=0.59$. 
It is apparent that the slow particles form clusters; moreover, 
there are more large clusters at $\phi=0.59$ than at $\phi=0.55$.    

The definition of the clusters shown in Fig.~\ref{slowclusters} is arbitrary. Alternative definitions results in different clusters.
For example, a more common definition uses the separation of the initial positions of slow particles corresponding to the 
first minimum of the respective pair correlation function,
\begin{equation}\label{pair}
g_{\alpha \beta}(r) 
= \frac{V}{N_\alpha (N_\beta - \delta_{\alpha \beta})} 
\left< \sum_{n}^{N_\alpha} \sum_{m \ne n}^{N_\beta} \delta (\mathbf{r} - \mathbf{r}_{nm}) \right>,
\end{equation}
where $V$ is the volume, $r=|\mathbf{r}|$, $\mathbf{r}_{nm} = \mathbf{r}_n - \mathbf{r}_m$, 
and the sums are over particles of $\alpha$ and $\beta$ type. Using such a definition we
find that the clusters span the entire simulation box. This 
is not surprising since, by definition, during time $\tau_\alpha$ on average 37\% of the particles
are slow. Thus within the first minimum of $g_{\alpha \beta}(r)$ of a given 
slow particle another slow particle is likely to be found.

To examine clusters of slow particles somewhat more quantitatively 
one can generalize the pair correlation functions, Eq.~\eqref{pair},
and define a correlation function involving slow particles only,
\begin{eqnarray}\label{4pointr}
G_4(r;t)  &=&  \frac{V}{\left< N_s(t) \right> (\left< N_s(t) \right> -1)}
\\ && \times \left< \sum_{n,m \ne n} w_n(t) w_m(t) 
\delta[\mathbf{r}-\mathbf{r}_{nm}(0) ]\right> \nonumber,
\end{eqnarray}
where microscopic single-particle overlap functions $w_n(t)$ select slow particles, 
and $\left< N_s(t) \right> $ is the average number of slow particles, see Eqs. (\ref{mof}-\ref{nosp}). 
Note that the summation in Eq. \eqref{4pointr} is over all, small and large, particles.
 
The function $G_4(r;t)$ is usually referred to as a four-point pair correlation function. 
Note that by definition, in the thermodynamic limit,
$G_4(r;t)\to 1$ as $r\to\infty$. By examining $G_4(r;t)-1$ we can examine the correlations
between slow particles. In particular, the spatial extent of these correlations manifests itself 
in a slower decay of $G_4(r;t)-1$ for large $r$. 

\begin{figure}
\includegraphics[width=3.2in]{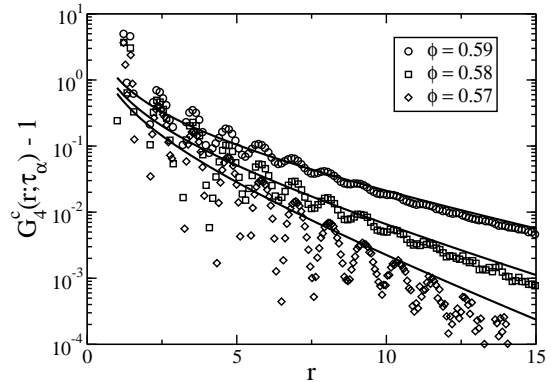}
\caption{\label{g4}The four-point correlation function $G_4^c(r;\tau_\alpha)-1$ for $\phi =$ 0.59, 0.58, and 0.57 
listed from top to bottom. 
The lines are proportional to $\exp[r/\xi(\tau_\alpha)]/r$ where $\xi(\tau_\alpha)$ are determined by
fits to $S_4(q;\tau_\alpha)$. }
\end{figure}

Investigation of the extent of the slow particles correlations through a direct analysis 
of $G_4(r)$ is complicated by finite size effects. In particular, in the finite system canonical ensemble 
the limiting large $r$ value of $G_4(r;t)$ differs from 1 by a term inversely proportional to the system size. 
To correct for this effect in a somewhat quantitative way
we determine the large $r$ limit of $G_4(r;t)$ by finding the average value of $G_4(r;t)-1$ from $r = 25.5$ to 
half the box length and then subtract this average from $G_4(r;t)$. The four-point function corrected in this way is 
denoted by $G_4^c(r;t)$. This function is shown in Fig.~\ref{g4}. We should emphasize that unlike in some other studies 
\cite{Doliwa2000,Lacevic2002,Rotman2010} we do not use this four-point function for a quantitative examination of
the slow particles correlations. For the latter task we found it more convenient to analyze the wave-vector 
dependent analog of $G_4(r;t)$.

Shown in Fig.~\ref{g4} is $G_4^c(r;\tau_\alpha)-1$ for $\phi = 0.57$, 0.58, and 0.59. The slower decay for
larger $\phi$ is evident, which indicates a larger correlation length associated with the average size of the
slow clusters with increasing $\phi$. 

To quantitatively determine the spatial extent of correlations of the slow particles 
we examine the $q$ dependence of the four-point structure factor, 
\begin{eqnarray}\label{4pointq}
S_4(q;t) &= & \frac{\left< N_s(t) \right> (\left<N_s(t) \right> - 1)}{N V} \tilde{G}_4(q;t) + \frac{\left<N_s(t)\right>}{N} 
\\ \nonumber 
&=& N^{-1} \left( \left< W(\mathbf{q},t)W(-\mathbf{q},t) \right> - \left|\left< W(\mathbf{q};t) \right>\right|^2 \right),
\end{eqnarray}
where $W(\mathbf{q};t)$ is the Fourier transform of the spatially resolved microscopic overlap function,
\begin{equation}
W(\mathbf{q};t) = \sum_n w_n(t) \exp[-i \mathbf{q} \cdot \mathbf{r}_n(0)]
\end{equation}
and $\tilde{G}_4(q;t)$ is the Fourier transform of $G_4(r;t)-1$. 

In the following two sub-sections we discuss two quantities that can be obtained from the four-point structure
factor: the dynamic susceptibility $\chi_4(t)$ and the dynamic correlation length $\xi(t)$.

\subsection{Dynamic Susceptibility $\chi_4(t)$}

The dynamic susceptibility, $\chi_4(t)$, is defined as the $q\to 0$ limit of the four-point structure factor,
\begin{equation}\label{chi4def}
\chi_4(t) = \lim_{q \rightarrow 0} S_4(q;t).
\end{equation} 
Since as $\mathbf{q}\to 0$, $W(\mathbf{q};t)\to \sum_n w_n(t) \equiv N_s(t)$ 
(note that in the preliminary report \cite{Flenner2010prl} we used $W(t)$ to denote the $\sum_n w_n(t)$) 
the dynamic susceptibility measures the ratio
of the fluctuation of the number of slow particles to the total number of particles, and we could formally write
\begin{equation}\label{chi4def1}
\chi_4(t) = N^{-1} \left( \left< N_s^2(t)\right> - \left< N_s(t) \right>^2 \right).
\end{equation}

It should be emphasized that the right-hand-side of Eq. \eqref{chi4def1} depends on the ensemble.
In the ensemble used in our study
the number of particles of both species are kept constant or, alternatively, the volume fraction $\phi$ and the concentration $c$
are kept constant. Thus, hereafter we will denote the right-hand-side of Eq. \eqref{chi4def1}
by $\chi_4(t)|_{\phi,c}$,
\begin{equation}\label{chi4pcdef}
\chi_4(t)|_{\phi,c} = N^{-1} \left( \left< N_s^2(t)\right> - \left< N_s(t) \right>^2 \right)
\end{equation}
where it is implicitly understood that the ensemble at the right-hand-side is the constant $\phi$ and $c$ simulational ensemble.
It should be noted that while $\chi_4(t)|_{\phi,c}$ can be easily calculated in a simulation, 
in order to determine the ensemble-independent susceptibility $\chi_4(t)$ one needs to perform a rather delicate 
extrapolation procedure $\lim_{q \rightarrow 0} S_4(q;t)$.

We note here that the difference between $\chi_4(t)|_{\phi,c}$ and $\chi_4(t)$ is the reciprocal space
manifestation of the finite size and ensemble dependencies of the large $r$ limit of the four-point correlation
function $G_4(r;t)$.  

Berthier \textit{et al.} \cite{Berthier2005} pointed out that 
the susceptibility $\chi_4(t)$ can be determined without extrapolating $S_4(q;t)$ by using
the formalism introduced in Ref. \cite{Lebowitz1967}. This procedure results in the following expression 
\begin{eqnarray}\label{chicorrection}
\chi_4(t) &=& \chi_4(t)|_{\phi,c} + \chi_\phi^2(t) H_1 + \chi_\phi(t) \chi_c(t) H_2 + \chi_c^2(t) H_3 \nonumber \\
&& + F_o^2(t) H_4 + F_o(t) \chi_\phi(t) H_5 + F_o(t) \chi_c(t) H_6,
\end{eqnarray} 
where $\chi_\phi(t) = \partial F_o(t)/\partial \phi$ and $\chi_c(t) = \partial F_o(t)/\partial c$.
The volume fraction dependent, but time independent, quantities $H_n$ are linear functions of the 
partial structure factors $S_{\alpha \beta}(q)$ extrapolated to $q=0$. Note that
we changed notation from previous work, Ref.~\cite{Flenner2010prl}, 
where we used $G_n$ instead of $H_n$.This was done to avoid confusion
with the four-point correlation function $G_4(r;t)$. 

In  Appendix \ref{xicalc} 
we present a derivation of Eq.~\ref{chicorrection}, give the explicit formulae for the quantities $H_n$, and
describe how these quantities were evaluated.   
In the same appendix we also describe an extrapolation procedure that confirmed the consistency of the definition
\eqref{chi4def} and the expression \eqref{chicorrection}.

It was further argued by Berthier \textit{et al.} \cite{Berthier2005} that 
Eq.~\eqref{chicorrection} could be used to establish an experimental lower bound
for $\chi_4(t)$. Since $\chi_4(t)|_{\phi,c}>0$ the correction terms at the right-hand-side of Eq.~\eqref{chicorrection}
constitute a lower bound for $\chi_4(t)$. Furthermore, since around the $\alpha$ relaxation time 
the first correction term is the dominant one, we could neglect all the other terms and thus arrive at
\begin{equation}
\chi_4(t) \ge \chi_\phi^2(t) H_1.
\end{equation} 
If it could be shown that the $\chi_\phi^2(t) H_1$ term dominates close to
the glass transition, one would have a simple approximation for the dynamic susceptibility. 
We start by examining the time and $\phi$ dependence of the terms on the right hand side of 
Eq.~\eqref{chicorrection} to examine this approximation in detail. 

\begin{figure}
\includegraphics[width=3.2in]{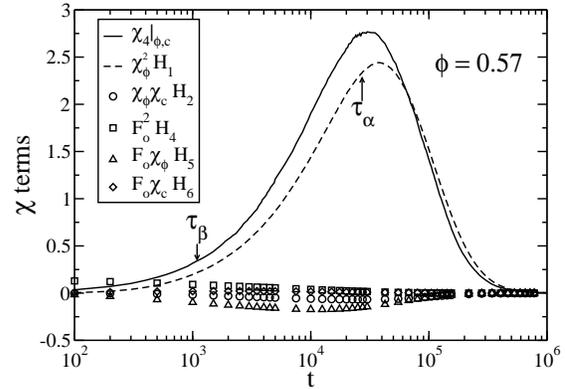}
\caption{\label{chitime}Time dependence of the terms that contribute to $\chi_4(t)$ for $\phi = 0.57$. The arrow
indicate the $\beta$ relaxation time $\tau_\beta$ and the $\alpha$ relaxation time $\tau_\alpha$.
}
\end{figure}

Shown in
Fig.~\ref{chitime} is $\chi_4(t)|_{\phi,c}$ and all the correction terms given in 
Eq.~(\ref{chicorrection}) for a representative volume fraction $\phi=0.57$. 
For very short times, the $F_o^2(t) H_4$
term is the largest, as it must be since it is the only term not equal to zero at $t=0$,
but it monotonically decays to zero. By the $\beta$ relaxation time, $\chi_4(t)|_{\phi,c}$
and the $\chi_\phi^2(t) H_1$ term are the largest, and by the $\alpha$ relaxation time they are around an order of magnitude larger
than the other terms.  For the volume fraction shown, these two terms are almost equal around $\tau_\alpha$, 
but the $\chi_\phi^2(t) H_1$ term becomes larger at later times.

\begin{figure}
\includegraphics[width=3.2in]{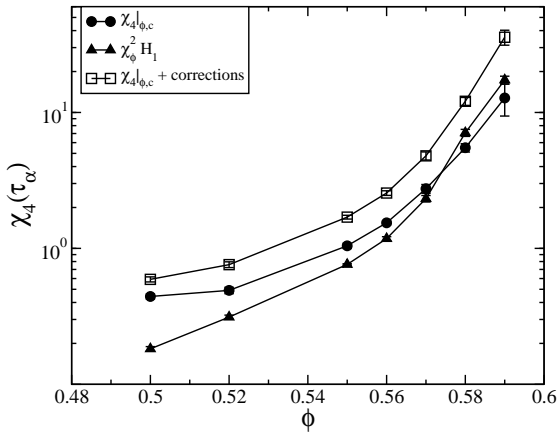}
\caption{\label{terms}Volume fraction dependence of 
the constant $\phi$ and $c$ part of the dynamic susceptibility, $\chi_4(\tau_\alpha)|_{\phi,c}$ (circles),
the dominant correction term, $\chi_\phi^2(\tau_\alpha) H_1$ (triangles), and the ensemble independent 
susceptibility $\chi_4(t)$ calculated from Eq. \eqref{chicorrection} (open squares).
}
\end{figure}

We note that with increasing volume fraction $\chi_\phi^2(\tau_\alpha) H_1$ grows faster than $\chi_4(\tau_\alpha)|_{\phi,c}$.
The $\chi_\phi^2(\tau_\alpha) H_1$ term becomes the dominant one 
for $\phi \ge 0.58$, Fig.~\ref{terms}. This is qualitatively consistent with result of Brambilla \textit{et al.} \cite{Brambilla2009}.
The quantitative difference between our Fig.~\ref{terms} and results shown in Fig. 3a
of Ref.~\cite{Brambilla2009} originates from the fact that Brambilla \textit{et al.}
systematically overestimated the isothermal compressibility which enters into
their correction term.

For our range of volume fractions we do not find that $\chi_4(\tau_\alpha)|_{\phi,c}$ is negligible compared
to $\chi_\phi^2(\tau_\alpha) H_1$. However, 
for volume fractions larger than the ones examined in this study, it is likely that $\chi_4(\tau_\alpha)$ is well
approximated by $\chi_\phi^2(\tau_\alpha) H_1$ term alone.

\begin{figure}
\includegraphics[width=3.2in]{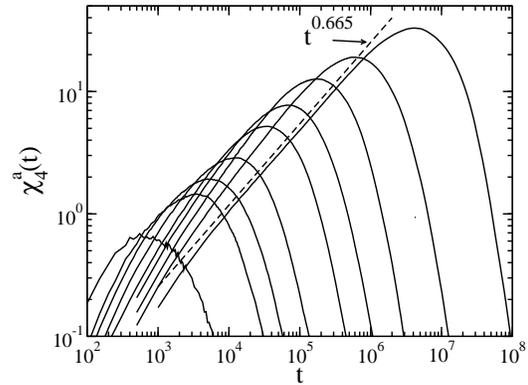}
\caption{\label{chia} The approximation of the dynamic susceptibility 
$\chi_4^a(t) = \chi_4(t)|_{\phi,c} + \chi_\phi^2(t) H_1$ for 
$\phi = 0.5$, 0.52, 0.54, 0.55, 0.56, 0.57, 0.575, 0.58, 0.585, and 0.59
listed from left to right. This approximation is accurate around $\tau_\alpha$,
i.e. around the peak shown in the figure, and becomes increasingly more accurate as $\phi$ increases.}
\end{figure}

As can be seen in Fig.~\ref{chitime}, around $\tau_\alpha$ a good approximation to $\chi_4(t)$ is 
\begin{equation}
\chi_4^a(t) = \chi_4(t)|_{\phi,c} + \chi_\phi^2(t) H_1. 
\end{equation}
The time dependence of this quantity is shown in 
Fig.~\ref{chia}. 
We observe that $\chi_4^a(t)$ grows with time, indicating an increase
of the overall strength of dynamic heterogeneity, until it reaches a peak that
occurs around $\tau_\alpha$ and then decreases 
to zero at later times. The decrease in $\chi_4^a(t)$ represents a diminishing of 
the overall strength of the heterogeneous dynamics but the length scale
associated with slow clusters do not have to follow the same trends.

Toninelli \textit{et al.}\ \cite{Toninelli2005} and Chandler \textit{et al.}\ \cite{Chandler2006} 
examined theoretical predictions for the time dependence of $\chi_4(t)$ and compared them with, \textit{inter alia}
particle-based simulations. It is not clear whether the latter comparisons 
were hindered by fact that global fluctuations were suppressed in simulations. However, in general, 
a common feature predicted by many theories is a power law growth of $\chi_4(t)$ while approaching the peak.
This fact prompted us to look for power laws in $\chi_4^a(t)$.  

For smaller volume fractions
we do not find any region of power law growth approaching $\tau_\alpha$. Around $\phi = 0.56$ there emerges a region where 
$\chi_4^a(t)$ appears to grow according to a power law, but the exponent in the power law depends on $\phi$. 
This is due to the two contributions to $\chi_4^a(t)$ having different magnitudes and time dependencies. 
For example, for $\phi=0.59$ we find that 
$\chi_4^a(t) \sim t^{0.665}$ in the $\alpha$ relaxation regime. This growth is due to a combination of 
$\chi_\phi^2(t) \sim t^{0.75}$ and $\chi_4|_{\phi,c} \sim t^{0.55}$ in the $\alpha$ relaxation time regime.  
This analysis suggests that the power law growth of $\chi_4^a(t)$ does not necessarily
have a deeper meaning, at least for volume fractions accessible in our study.

As we remarked above, we expect that for sufficiently high volume fractions the growth of $\chi_4(t)$ 
can be obtained from experiments using the $\chi_\phi^2(t) H_1$ correction term as an approximation for $\chi_4(t)$.
Moreover, if time-temperature superposition holds, then the growth of  $\chi_\phi^2(t)$ is related to the
growth of the $\alpha$ relaxation time. 
Below, we investigate the consequencies of this idea.

For hard sphere systems, time-temperature superposition is replaced by time-volume fraction superposition. 
Specifically, the statement is that
$F_o(t/\tau_\alpha)$ overlaps in the $\alpha$ relaxation regime when plotted
for different $\phi$. We find good overlap for $\phi \ge 0.58$, although
we do observe small systematic deviations. Moreover, we find that 
$F_o(t/\tau_\alpha)$ is well described by a stretched exponential $A e^{-(t/\tau_\alpha)^\beta}$ 
(see Fig.~\ref{fig:Fo}). Thus, ignoring the weak volume fraction dependence of $A$ and $\beta$
the first correction term in Eq.~\eqref{chicorrection} at $\tau_\alpha$ is given by
\begin{equation}
\chi_\phi^2(\tau_\alpha) H_1 = \left( \frac{2 A \beta}{(d_1^3 + d_2^3)} \right)^2 
\left( \frac{\partial \ln(\tau_\alpha)}{\partial \ln(\phi)} \right)^2 e^{-2} H_1.
\end{equation}
Recall that this is the largest term for $\phi \ge 0.58$. Since liquid structure is weakly volume-fraction dependent,  
$H_1$ changes slowly with $\phi$. Notice that $A$, $\beta$ and $H_1$ are all less
than one or equal to one for all $\phi$, and are all less than one for $\phi \ge 0.58$. Note also
that $2/(d_1^3 + d_2^3) = 0.53419/d_1^3$. Thus, the coefficient multiplying 
$[\partial \ln(\tau_\alpha)/\partial \ln(\phi)]^2$ is less than one at all $\phi$ and is 
very weakly $\phi$ dependent.  Consequently, $\chi_4(t)$ behaves as
$[\partial \ln(\tau_\alpha)/\partial \ln(\phi)]^2$ when the $\chi_\phi^2$ term is dominant. 
Finally, since at the largest volume fractions $\tau_\alpha = \tau_\infty \exp[B/(\phi_0 - \phi)^2]$
provides a good fir to our data, then
these arguments indicate that $\chi_4(\tau_\alpha) \sim \phi^2 (\phi_0 - \phi)^{-6}$ close to 
$\phi_0$. We find that this indeed provides a good description of our results for the dynamic susceptibility (See Fig.~\ref{chixi}). 

\subsection{Dynamic Correlation Length $\xi(t)$}
To define the dynamic correlation length $\xi(t)$ we need to examine the long wavelength (small wave-vector)
behavior of the four-point structure factor. Specifically, we use the following definition of 
the dynamic correlation length:
\begin{equation}\label{xidef}
\xi^2(t) = \lim_{q\to 0} q^{-2} \left(\frac{\lim_{k\to 0} S_4(k;t)}{S_4(q;t)}-1\right).
\end{equation}
Definition \eqref{xidef} is consistent with asymptotic small wave-vector Ornstein-Zernicke form of the 
four-point structure factor, 
\begin{equation}
S_4(q;t) \approx \frac{\chi_4(t)}{1+\xi(t)^2q^2}\; \mathrm{as}\; q\to 0.
\end{equation} 
We should note that formally definition \eqref{xidef} is equivalent to defining $\xi(t)$ as the second moment of $G_4(r;t)-1$
divided by the zeroth moment of $G_4(r;t)-1$. However, we found that the finite size and ensemble effects are easier
to account for in the reciprocal space and therefore we used definition \eqref{xidef}. 

\begin{figure}
\includegraphics[width=3.2in]{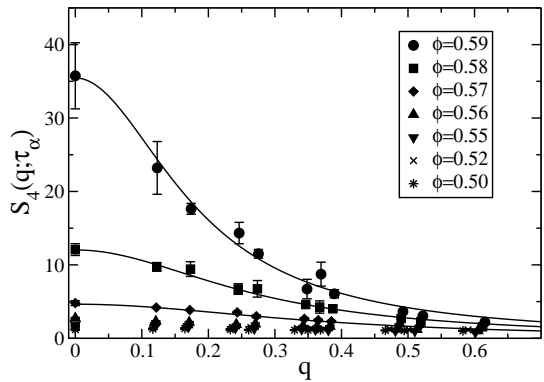}
\caption{\label{s4} Dynamic structure factor $S_4(q;\tau_\alpha)$ for $\phi=$ 0.59, 0.58, 0.57, 0.56, 0.55, 0.55, 0.52, and 0.50.
The values at $q=0$ were calculated using Eq.~\eqref{chicorrection}.
The lines are Ornstein-Zernicke fits.}
\end{figure}

To obtain reliable results for $\chi_4(t)$ and $\xi(t)$ we fitted the simulation results to several different
functional forms (see Appendix \ref{xicalc} for details). We determined that the best procedure is to 
fit $S_4(q;t)$, including as $q=0$ value the right-hand-side of Eq.~\eqref{chicorrection}, to the Ornstein-Zernicke
form while restricting the fitting range to $q < 1.5/\xi(t)$. Such fits at $\tau_\alpha$ are shown in Fig.~\ref{s4},
for volume fractions $\phi = 0.59$, 0.58, 0.57, 0.56, 0.55, 0.52, and 0.50, calculated using the $80\, 000$ particle
simulations (note the the system size dictates the smallest non-zero wave-vector).

\begin{figure}
\includegraphics[width=3.2in]{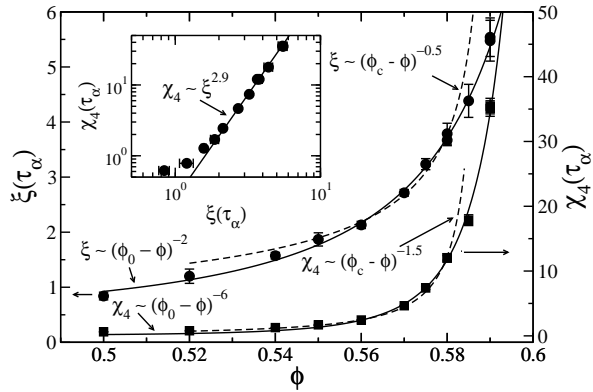}
\caption{\label{chixi}The dynamic correlation length $\xi(\tau_\alpha)$ versus 
$\phi$ (left axis) and the dynamic susceptibility $\chi_4(\tau_\alpha)$ versus $\phi$ (right axis). 
The solid line through $\xi(\tau_\alpha)$ is a fit to $\xi(\tau_\alpha) = \xi_0 + A(\phi_0 - \phi)^{-2}$
where $\phi_0 = 0.635$. The dashed line through $\xi(\tau_\alpha)$ is a mode-coupling like
fit $\xi(\tau_\alpha) \sim (\phi_{c} - \phi)^{-\gamma_\xi}$ where $\phi_{c} = 0.59$ is fixed and
we obtain $\gamma_\xi = 0.5 \pm 0.1$. In the inset we show $\chi_4(\tau_\alpha)$ versus $\xi(\tau_\alpha)$,
and we find that $\chi_4(\tau_\alpha) \sim \xi(\tau_\alpha)^{2.9}$ for $\phi \ge 0.55$. }
\end{figure}

Shown in Fig.~\ref{chixi} is the volume fraction dependence of the resulting dynamic susceptibility and correlation length.
Note that in this figure we also included results obtained applying the same procedure to data obtained from $10\, 000$ particle
simulations. The consistency of both sets of results indicates that our procedure can be used to determine the 
dynamic susceptibility and correlation length using moderately large systems.
 
We find that the volume fraction dependence of $\xi(\tau_\alpha)$ can be well described 
by many different fit functions (the results
of the mode-coupling like fits are described in detail in Appendix \ref{fits}). To be
consistent with previous fits to $\tau_\alpha$ and $D$, we fix $\phi_{c} = 0.59$ and fit $\xi(\tau_\alpha)$ 
to a mode-coupling like power law, $\xi(\tau_\alpha) \sim (\phi_{c} - \phi)^{-\gamma_\xi}$. This 
results in $\gamma_{\xi} =  0.5 \pm 0.1$. The corresponding fit is shown as the dashed line in Fig.~\ref{chixi}. 
The value of $\gamma_{\xi}$ obtained from the fit does not agree with the inhomogeneous mode-coupling prediction of 
$\gamma_{\xi} = 0.25$ \cite{Biroli2006,Szamel2010}.

Next we fit $\xi(\tau_\alpha)$ to $\xi_0 + C(\phi_0 - \phi)^{-2}$
over the whole range of $\phi$, which gives $\phi_0 = 0.0635 \pm 0.004$
and $\xi_0 = 0.37 \pm 0.1$. The corresponding fit  
is shown as the solid line in Fig.~\ref{chixi}, and provides an accurate description of $\xi(\tau_\alpha)$
for every volume fraction examined in this work. 
Note that the same $\phi_0 = 0.635$ was obtained from fits of the $\alpha$ relaxation time to the formula 
suggested by Berthier and Witten \cite{Berthier2009}, 
$\tau_\alpha = \tau_\infty \exp[B (\phi_0 - \phi)^{-2})]$ (see Appendix \ref{fits}). This observation suggests that 
the following correlation between the $\alpha$ relaxation time and the length, 
$\tau_\alpha = \tau_0 \exp[k \xi(\tau_\alpha)]$. We discuss this relationship in Sec.~\ref{lengthdynamics}. 

We now look at the scaling relationship between the dynamic susceptibility and the length,
$\chi_4(\tau_\alpha) \sim \xi(\tau_\alpha)^{2-\eta}$. For compact clusters it is expected that
$2-\eta = d$ where $d$ is the spatial dimension. Shown in the inset to Fig.~\ref{chixi} is the scaling
fit for $\phi \ge 0.56$. We obtain $2-\eta = 2.9 \pm 0.1$, which indeed suggests compact clusters.
The comparison of this result with the inhomogeneous mode-coupling theory prediction \cite{Biroli2006,Szamel2010} 
is a little involved. The theory analyzes a three-point susceptibility and finds that in the $\alpha$ relaxation
regime $\lim_{\mathbf{q}\to 0} \chi_3(\mathbf{q};\tau_\alpha) \sim \xi(\tau_\alpha)^4$. Field-theoretical considerations 
\cite{Berthier2007p2,Berthier2007p3} indicate that the dynamic susceptibility is a quadratic function of the 
three-point susceptibility. A combination of both results would suggest a prediction 
$\chi_4(\tau_\alpha) \sim \xi(\tau_\alpha)^8$ which is clearly well outside the simulational result. 

Since we find that $\xi(\tau_\alpha) = \xi_0 + C(\phi_0 - \phi)^{-2}$ provides a good description of all the data and
that $\chi_4(\tau_\alpha) \approx a_4 \xi(\tau_\alpha)^3$ for $\phi \ge 0.55$, we show  
$a_4 [\xi_0 + C(\phi_0 - \phi)^{-2}]^3$ as the solid line through $\chi_4(\tau_\alpha)$
in Fig.~\ref{chixi}. Note that this is not an independent fit. However, it describes the data fairly
well over the whole range of studied volume fractions. Moreover, it is consistent with the limiting behavior
$\chi_4(\tau_\alpha) \sim \phi^2 (\phi_0 - \phi)^{-6}$ obtained from the first correction in Eq.~\eqref{chicorrection}.

In Fig.~\ref{chixi} we also show, as a dashed line, the third power of the
mode coupling fit. As expected, it gives a reasonable description of the data between
$0.55 \le \phi \le 0.58$.

\begin{figure}
\includegraphics[width=3.2in]{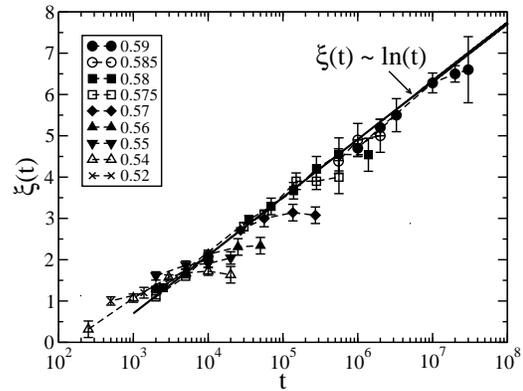}
\caption{\label{xitime}The dynamic correlation length versus time for $\phi \ge 0.52$. The
correlation length appears to follow a universal curve until it reaches a volume fraction dependent maximum value, 
and then it stays approximately constant at later times. The straight 
line is a fit to the data, $\xi(t) = a\ln(b t)$.}
\end{figure}

We now examine the time dependence of the dynamic correlation length. 
Shown in Fig.~\ref{xitime} is $\xi(t)$ versus time for different $\phi$. For all $\phi$, the dynamic correlation
length grows with time and then plateaus 
at later times, and remains constant up to the maximum time at which we can evaluate it. 
We cannot accurately calculate $\xi(t)$ for $t \gtrsim 10 \tau_\alpha$
because there are few slow particles at such long times. Thus, we do not know
the fate $\xi$ at later times. 

We note that one previous simulational investigation of the time dependence of the dynamic correlation length
resulted in length whose time dependence roughly follows that of the dynamic susceptibility \cite{Lacevic2003}.
A different study found a monotonically increasing dynamic correlation length \cite{Toninelli2005}.
On the other hand, an earlier study \cite{Doliwa2000}, which used a somewhat different definition of 
the dynamic correlation length, found that the length was increasing with time but then plateaued after $\tau_\alpha$.
A similar behavior was found in a very recent study of a two-dimensional lattice gas glassy system \cite{Rotman2010}.
The two latter results are (at least qualitatively) consistent with our findings. 

We remark that a plateau in the time dependence of a characteristic dynamic length is predicted 
by the inhomogeneous mode-coupling theory \cite{Biroli2006,Szamel2010}. However, the plateau predicted by this theory
occurs around the $\beta$ relaxation time, and not after the $\alpha$ relaxation time as seen here.

There are two somewhat surprising features in the results shown in Fig.~\ref{xitime}.
First, for $\phi \ge 0.56$ and between $\tau_\beta$ and $\tau_\alpha$, 
the dynamic correlation length is independent of the volume fraction and only depends on time. 
We don't have sufficient data for times smaller than the $\beta$ relaxation time, but we expect that
this universal behavior breaks down at short times. The correlation length 
appears to follow a master curve until it reaches a volume fraction dependent 
asymptotic value, which we will refer to $\xi_{\mathrm{max}}$. 
We find that this master cure can be well described
by $\xi(t) = a\ln(b t)$, and this fit is shown as a solid line in Fig.~\ref{xitime}.

Second, we find that the time at which $\xi(t)$ saturates, which we denote as $\tau_{\mathrm{max}}$,
exceeds the $\alpha$ relaxation time and the time at which $\chi_4(t)$ peaks, $\tau_{\mathrm{peak}}$
(we find that $\tau_\alpha$ and $\tau_{\mathrm{peak}}$ have the same volume fraction dependence). 
Thus, the dynamic correlation length seems to be growing further while the overall strength of dynamic
heterogeneity, measured by the susceptibility, is decreasing.  This 
somewhat surprising finding means that, while at longer time scales there are few
slow particles, the characteristic size of the clusters of these particles seems to be constant (at least up to 
$10 \tau_\alpha$).

\begin{figure}
\includegraphics[width=3.2in]{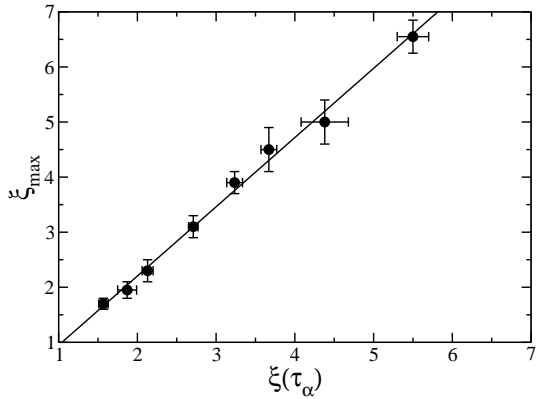}
\caption{\label{xi2time}The plateau value $\xi_{\mathrm{max}}$ versus $\xi(\tau_\alpha)$. The solid
line is a linear fit.}
\end{figure}

Since $\tau_{\mathrm{max}}$ exceeds $\tau_\alpha$, it is obvious that 
$\xi_{\mathrm{max}}$ exceeds $\xi(\tau_\alpha)$. Interestingly, there seems to be a linear relationship between
these two lengths, 
Fig.~\ref{xi2time}. Combining 
the linear relationship between $\xi_{\mathrm{max}}$ and $\xi(\tau_\alpha)$ with the previous 
observation that $\xi(t) = a \ln(bt)$, we see that $\tau_{\mathrm{max}} \sim \tau_\alpha^\epsilon$ and 
through the fits of $\xi_{\mathrm{max}}$ we determine $\epsilon = 1.3 \pm 0.1$.  

The time dependence of the dynamic correlation length shown in Fig.~\ref{xitime} suggests that possibly we should 
focus more on the plateau value of the dynamic correlation length, $\xi_{\mathrm{max}}$, than on the length at 
the $\alpha$ relaxation time, $\xi(\tau_\alpha)$. This suggestion is left for future investigation.

\section{correlation length and average dynamics}\label{lengthdynamics}

In this section we examine the relationships between the dynamic correlation length $\xi(\tau_\alpha)$
and the two simplest quantities characterizing the average dynamics, the relaxation time,
$\tau_\alpha$, and the self-diffusion coefficient, $D$. We note that 
most theoretical descriptions of glassy dynamics focus on the temperature dependence of the dynamics. Consequently, 
relationships between $\xi(\tau_\alpha)$, $\tau_\alpha$ and $D$ involve temperature. 
However, temperature is not a relevant control variable for our hard sphere system.
Instead, the volume fraction is the usual control parameter and, therefore, in formulae discussed below we omitted temperature.

We begin by examining the relationship between $\xi(\tau_\alpha)$ and $\tau_\alpha$. 
The mode-coupling theory predicts a power law $\tau_\alpha \sim \xi(\tau_\alpha)^z$. In contrast, the
Adam-Gibbs \cite{Adam1965} and Random-First-Order-Transition theories \cite{Kirkpatrick1989} predict
an exponential dependence of the relaxation time on a correlation length, $\xi_s$, describing the size of cooperatively 
rearranging regions, $\tau_\alpha \sim \exp(\xi_s^\theta)$ where $\theta = 3$ in the Adam-Gibbs theory 
and is a parameter in the Random-First-Order-Transition theory. While it
is currently unclear if our dynamic correlation length is the same as the correlation length 
in Adam-Gibbs and  Random-First-Order-Transition theories (in particular our length depends on time whereas
$\xi_s$ does not have an obvious time dependence) 
we examine relationships between $\xi$ and $\tau_\alpha$ suggested by those theories.

\begin{figure}
\includegraphics[width=3.2in]{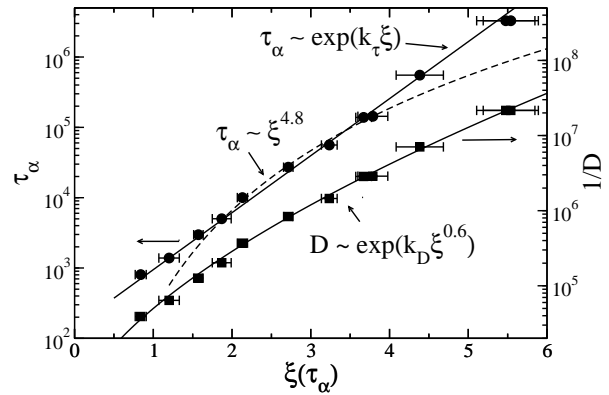}
\caption{\label{tauDxi}The alpha relaxation time $\tau_\alpha$ and the diffusion coefficient $D$ 
as a function of $\xi(\tau_\alpha)$. The solid straight line is an exponential fits to the relaxation time data 
$\tau_\alpha \sim \exp[\xi(\tau_\alpha)]$ and the solid curved line is a fit $1/D \sim \exp[\xi(\tau_\alpha)^\theta]$ 
with $\theta = 0.6$. The dashed line is a mode-coupling theory like fit to $\tau_\alpha \sim \xi(\tau_\alpha)^z$
with $z = 4.8$.}
\end{figure}

We find that a power law describes well the correlation between $\xi(\tau_\alpha)$ and $\tau_\alpha$ over the mode-coupling regime 
with $z=4.8\pm0.3$, see Fig.~\ref{tauDxi}. This exponent disagrees with the
inhomogeneous mode-coupling theory prediction of $z\approx 10$ \cite{Biroli2006,Szamel2010} and with 
some of the previous simulational studies 
\cite{Lacevic2003,Stein2008}, but it is consistent with other previous simulational investigations 
\cite{Whitelam2004,Flenner2009,Karmakar2009} (note that in majority of earlier
studies the inverses of the exponent $z$ was given).

We find, however, that an exponential dependence
of $\tau_\alpha$ on $\xi(\tau_\alpha)$ provides a better description of the data over a larger range of volume fractions. 
A fit to $\tau_\alpha = \tau_0 \exp(k_\tau \xi(\tau_\alpha)^\theta)$ gives $\theta = 1.1 \pm 0.2$. Thus, 
we fix $\theta = 1.0$ and fit $\tau_\alpha = \tau_0 \exp(k_\tau \xi(\tau_\alpha))$, which is shown
as a solid line in Fig.~\ref{tauDxi}. Note that the quality of the latter fit is fully consistent with the 
fact that independent fits $\tau_\alpha = \tau_\infty \exp[B(\phi_0 - \phi)^{-2}]$ and 
$\xi(\tau_\alpha) = \xi_0 + C(\phi_0 - \phi)^{-2}$ result in the same value $\phi_0 = 0.635$ (see Appendix \ref{fits}).

In Fig.~\ref{tauDxi} we also show $1/D$ versus $\xi(\tau_\alpha)$. The results do not seem to follow a 
straight line and thus we do not find that $D \sim \exp(-k_D \xi(\tau_\alpha))$, but rather 
$D = D_0 \exp(-k_D \xi(\tau_\alpha)^\theta)$ where $\theta = 0.61 \pm 0.04$. The latter fit is shown as a solid line in the figure. 
Again, we note that the quality of the self-diffusion coefficient fit is quite good. However,
we shall also note that a combination of both correlations $\tau_\alpha = \tau_0 \exp(k_\tau \xi(\tau_\alpha))$
and $D = D_0 \exp(-k_D \xi(\tau_\alpha)^\theta)$ is, strictly speaking, not compatible with a power-law relationship between the 
self-diffusion coefficient and the relaxation time discussed in Sec.~\ref{dynamics}

Finally, we briefly mention two other, different investigations that analyzed somewhat different 
characteristic dynamic lengths. 

Saltzman and Schweizer \cite{Saltzman2006a,Saltzman2006b,Saltzman2008} investigated a characteristic length
associated with the onset of Fickian diffusion. They showed that this length, $\xi_D$, depends logarithmically on the 
relaxation time, $\xi_D \sim \ln(\tau_\alpha)$. This result is consistent with our relation between the 
dynamic correlation length at the $\alpha$ relaxation time and the $\alpha$ relaxation time. 

A similar crossover length was examined in models of facilitated dynamics \cite{Berthier2005a}.
Berthier \textit{et al.}\ defined a length scale associated with the onset of Fickian diffusion as 
$\ell^* \sim \sqrt{D \tau_\alpha}$. They noted that  
$\ell^* \sim \tau_\alpha^{(1-\sigma)/2}$ where $\sigma$ is the dynamic exponent describing the violation
of the Stokes-Einstein relation. Thus, for our system one 
would expect $\ell^* \sim \tau_\alpha^{0.2}$. We do find that for our system the relation 
$\sqrt{D \tau_\alpha} \sim \tau_\alpha^{0.2}$ is obeyed for large $\tau_\alpha$.
However, the length $\ell^*$ has different volume fraction (or relaxation time) dependence from our 
dynamic correlation length $\xi(\tau_\alpha)$. 
This is qualitatively consistent with the fact that the analysis of facilitated models suggests that 
the length $\ell^*$ is actually different from a dynamic heterogeneity length $\xi$. For example, for the so-called East model,
one finds the following relation between the dynamic heterogeneity length and the $\alpha$ relaxation
time, $\xi \sim \exp\{\sqrt{\ln(\tau_\alpha)}\}$  \cite{GarrahanPC}. This relation, for our range of
correlation lengths and relaxation times, provides as good fit to our data as the logarithmic relation
$\xi(\tau_\alpha) \sim \ln(\tau_\alpha)$ discussed above.

\section{Discussion}\label{conclusions}

We presented a new computational method to calculate a dynamic correlation length $\xi(t)$ characterizing the spatial
extent of heterogeneous dynamics on time scale $t$ and used this new method to obtain a number of results pertaining to
dynamic heterogeneity. 

Our method combines direct simulational evaluation of the four-point structure factor $S_4(q;t)$ for non-zero
wave-vectors with an independent calculation of the dynamic susceptibility
$\chi_4(t)$ that accounts for fluctuations suppressed in the simulational 
ensemble via procedures derived by Lebowitz \textit{et al.}\ \cite{Lebowitz1967}. Using the independently obtained 
dynamic susceptibility as the $q\to 0$ limit of $S_4(q;t)$ facilitates analyzing the small $q$ behavior of the four-point 
structure factor. We found that an Ornstein-Zernicke fits worked well if we restricted our fits such that $q < 1.5/\xi(t)$. 
This procedure allows one to evaluate the dynamic correlation length from simulations of moderately large systems. We also found 
that the calculation of $\xi(t)$ from the direct space four-point correlation function $G_4(r;t)$ is difficult due to the  
difficult to account for finite size and ensemble dependencies. 

We studied the volume fraction and the  time dependence of the dynamic correlation length. We also explored
relationships between the length, the dynamic susceptibility, and quantities characterizing the average 
dynamics, the $\alpha$ relaxation time and the self-diffusion coefficient.

First, we found that $\xi(\tau_\alpha) \sim (\phi_0 - \phi)^{-2}$ provides a good description of the data. 
We note, however, $\xi(\tau_\alpha)$ can also be fitted by other functions. We also found that
mode-coupling like power law  $\xi(\tau_\alpha) \sim (\phi_c - \phi)^{-\gamma_\xi}$ provides 
a good description of $\xi(\tau_\alpha)$ for the mode-coupling theory 
range of volume fractions, but the exponent $\gamma_\xi = 0.5$ differs from the inhomogeneous
mode-coupling theory prediction of $\gamma_\xi = 0.25$ \cite{Biroli2006,Szamel2010}. 

Next, we studied the time dependence of $\chi_4(t)$ and $\xi(t)$. While we did find a power law dependence on time, 
$\chi_4(t) \sim t^c$, for times around the $\alpha$ relaxation time, the exponent $c$ was volume fraction dependent 
and decreased with increasing $\phi$. Surprisingly, we found that for
a range of times between the $\beta$ and $\alpha$ relaxation times the dynamic correlation length 
followed a master curve independent of $\phi$ until it reached a volume fraction dependent plateau value.
The dependence of $\xi(t)$ on time could be fitted with a simple $\xi(t) \sim \ln(t)$ relation. 
The plateau value, $\xi_{\mathrm{max}}$, was reached at a  characteristic time, $\tau_{\mathrm{max}}$. We 
found that $\tau_{\mathrm{max}}$ exceeds and grows faster with increasing volume fraction than the $\alpha$ relaxation time. 

We examined the correlations between $\xi(\tau_\alpha)$, $\tau_\alpha$, and $D$, and we found that mode-coupling
like power law fits provide a good description of the data for $0.55 \le \phi \le 0.58$. We found 
deviations from these fits as $\phi_{c} = 0.59$ is approached. While the mode-coupling exponents for 
$\tau_\alpha$ and $D$ agree reasonably 
well with the mode-coupling predictions, the exponents for $\xi(\tau_\alpha)$ and $\chi_4(\tau_\alpha)$
do not. We also find that $\chi_4(\tau_\alpha) \sim \xi(\tau_\alpha)^3$, which does not agree with the
inhomogeneous mode coupling prediction. 

Finally, we found that an exponential dependencies of the $\alpha$ relaxation time and the self-diffusion coefficient
on the dynamic correlation length evaluated at the $\alpha$ relaxation time,
$\tau_\alpha \sim \exp(\xi(\tau_\alpha))$ and $D \sim \exp(-\xi(\tau_\alpha)^{0.6})$, describe our data well. This is
consistent with the spirit of Adam-Gibbs and Random-First-Order-Transition theories. The values of the scaling exponents are
inconsistent with the traditional Adam-Gibbs picture where the relaxation, either $\tau_\alpha$ or $1/D$,
behaves as $\exp(\xi_s^3)$. Note, however, that Adam-Gibbs and  Random-First-Order-Transition theories are formulated
in terms of the characteristic size of dynamically correlated regions, $\xi_s$ whereas we calculated and examined
the dynamic correlation length. Further work is required to clarify the connection between these lengths. 

\section{Acknowledgments}
We gratefully acknowledge the support of NSF Grant No.\ CHE 0909676.

\appendix
\section{Characteristic Volume Fractions}\label{fits}
To test theories of the glass transition it is common to fit experimental and simulation data to different functions.
The quality of the fits can vary depending on the range used for the fits and the proximity to any singularity 
implied by the fitting function. In this section we examine some commonly used fitting functions
to various $\phi$ dependent quantities,
a mode-coupling like fit $(\phi_{c} - \phi)^{-\gamma}$, a Vogel-Fucher-Tamman (VFT) like form 
$\exp[A (\phi_{VF}-\phi)^{-1}]$ and a form suggested by Berthier and Witten (BW), $\exp[B (\phi_0 - \phi)^{-2}]$
\cite{Berthier2009}.
The mode-coupling fits are used to determine a range of volume fractions where the mode-coupling like 
power laws provide a good description of the data. We use this range of $\phi$ to compare our simulation results
to the predictions of the mode-coupling theory and the inhomogeneous mode-coupling theory. Outside of this
range we do not expect the mode-coupling theory to provide a very good description of the dynamics. 
The goal of this appendix is to examine our results and previous arguments in the literature to 
find the best unified description of the data. To achieve this goal, we not only examine our data, but
also use results from earlier investigations\cite{Odriozola2010,Berthier2009}.

First we examine mode-coupling like fits to the $\alpha$ relaxation time $\tau_\alpha$ and the self-diffusion
coefficient $D$.  The mode-coupling theory predicts a power law divergence of $\tau_\alpha$ and
a power law vanishing of $D$ with the same exponent $\gamma$. It has been found in several 
numerical studies of the mode-coupling theory that $\gamma \approx 2.46$ \cite{Flenner2005mc,Szamel2010}. 
One of the difficulties in performing these fits is that, while
the mode-coupling theory predicts that Stokes-Einstein relation is obeyed, $\tau_\alpha D = \mathrm{const.}$, 
\cite{Flenner2005mc}, this relation is violated in simulations and experiments. Importantly, in most simulations the violation
of the Stokes-Einstein relation is apparent already in the regime in which power law fits are applicable. Thus, the best
one can do is to use different exponents for $\tau_\alpha$ and $D$ and the same mode-coupling transition 
volume fraction or temperature. A seemingly worse alternative is to force the same exponent and obtain two different
mode coupling transition points. 

We fit $\tau_\alpha$, $1/D$, $\xi(\tau_\alpha)$, and $\chi_4(\tau_\alpha)$
to power laws of the form $a(\phi_{c} - \phi)^{-\gamma}$ for $0.55 \le \phi \le 0.575$ and $0.55 \le \phi \le 0.58$. The results
are summarized in Table \ref{mcfits}. We find that $\phi_{c}$ varies from $0.58$ to 0.61, but
a consistent value is around $\phi_{c} \approx 0.59$. Since $\phi_{c} = 0.59$ is consistent
with our results, has been used in the literature previously \cite{Brambilla2009}, and coincides with the
onset of "hopping" like motion observed in Sec.~\ref{dynamics}, 
we fix $\phi_{c} = 0.59$
in this work. We also identify $0.55 \le \phi \le 0.58$ as the mode-coupling regime, but this should
be considered as only an approximate regime where the mode-coupling theory provides a reasonable 
description of the data.

Having chosen a common value for the mode-coupling transition volume fraction we redo the fits for 
$1/D$, $\tau_\alpha$, $\chi_4(\tau_\alpha)$, and $\xi(\tau_\alpha)$ keeping $\phi_c = 0.59$ fixed.
The fit parameters are the bottom set in Table \ref{mcfits}. 

\begin{table}
\caption{\label{mcfits}Fits to a mode-coupling like power law over different ranges of volume functions.
The number in parenthesis represents the uncertainty in the last digit. The bottom set of data is with 
$\phi_c = 0.59$ fixed.}
\begin{ruledtabular}
\begin{tabular}{|c|d|d|c|}
variable & \phi_{c} & \gamma & fit range\\ \hline
$\tau_\alpha$ & 0.5874(4) & 2.21(6) & $0.55 \le \phi \le 0.575$ \\
$D$ & 0.5911(7) & 2.13(6) & $0.55 \le \phi \le 0.575$ \\
$\chi_4(\tau_\alpha)$ & 0.58(4) & 1.3(2) & $0.55 \le \phi \le 0.575$ \\
$\xi(\tau_\alpha)$ & 0.586(4) & 0.47(9) & $0.55 \le \phi \le 0.575$ \\ 
\hline
$\tau_\alpha$ & 0.5901(6) & 2.47(6) & $0.55 \le \phi \le 0.58$ \\
$D$ & 0.5950(4) & 2.42(5) & $0.55 \le \phi \le 0.58$ \\
$\chi_4(\tau_\alpha)$ & 0.593(3) & 1.7(2) & $0.55 \le \phi \le 0.58$ \\
$\xi(\tau_\alpha)$ & 0.61(2) & 0.9(4) & $0.55 \le \phi \le 0.58$ \\
\hline
$\tau_\alpha$ & 0.59 & 2.43(1) & $0.55 \le \phi \le 0.58$ \\
$D$ & 0.59 & 1.94(3) & $0.55 \le \phi \le 0.58$ \\
$\chi_4(\tau_\alpha)$ & 0.59 & 1.46(4) & $0.55 \le \phi \le 0.58$ \\
$\xi(\tau_\alpha)$ & 0.59 & 0.50(3) & $0.55 \le \phi \le 0.58$ \\
\end{tabular}
\end{ruledtabular}
\end{table}

We now look at the fate of the system beyond the mode-coupling regime. To this end we examine
the results of fits to a VFT like functions and a BW like function for $\tau_\alpha$. We do not fit
other variables since we do are not sure whether the same functions can be used. Note,
however, that $\xi(\tau_\alpha)$ is closely tied to $\tau_\alpha$, thus we expect that fits 
to $\xi(\tau_\alpha)$ result in the same conclusions. 

\begin{figure}
\includegraphics[width=3.2in]{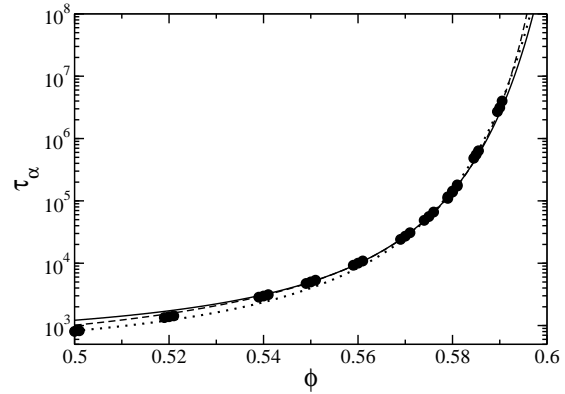}
\caption{\label{taufits}Various fits to $\tau_\alpha$ described in the text. The dotted 
line is a combination of the fit $\xi(\tau_\alpha) \sim (\phi_0 - \phi)^{-2}$ and the correlation 
$\tau_\alpha \sim \exp[k \xi(\tau_\alpha)]$. 
The dashed line is the Vogel-Fucher-Tamman fit and the solid line is a fit to 
$\tau_\alpha \sim \exp[B(\phi_0 - \phi)^{-2}]$.}
\end{figure}

A fit to $\ln(\tau_\alpha) = \ln(\tau_V) +  A(\phi_V - \phi)^{-1}$ gives 
$\phi_V = 0.6122 \pm 0.0005$, $\tau_V = 140 \pm 7$ and $A = 0.222 \pm 0.005$ where
we fit $0.55 \le \phi \le 0.5905$. Next we fit $\ln(\tau_\alpha) = \ln(\tau_0) + B(\phi_0 - \phi)^{-2}$, which
gives $\tau_0 = 456 \pm 37$, $B = 0.017 \pm 0.001$, and $\phi_0 = 0.635 \pm 0.002$.  Shown in
Fig.~\ref{taufits} are these fits along with the results obtained by fitting $\xi(\tau_\alpha) = \xi_0 + C(\phi_0 - \phi)^{-2}$
and then using $\tau_\alpha \sim \exp(k \xi)$, which we refer to as the correlation length fit. 
The VFT fit is the best fit over the largest range of $\phi$, thus one would choose
this fit based on the fit quality alone. However, the VFT fit results in a critical volume fraction that 
appears to be too small when compared with earlier results from the literature. 

One result is the dynamic scaling argument of Berthier and Witten \cite{Berthier2009}
who found $\phi_0 = 0.635 \pm 0.005$ and $\tau_\alpha \sim \exp[B(\phi_o - \phi)^{-\delta}]$
with $\delta = 2.2 \pm 0.2$. This is in remarkable agreement with our fits to 
$\tau_\alpha$ and $\xi(\tau_\alpha)$. Another is the
work of Odriozola and Berthier \cite{Odriozola2010} who found no evidence of a thermodynamic 
transition for $\phi < 0.63$
by utilizing a replica exchange Monte Carlo algorithm to examine the equation of state. 
Finally, we find that $\tau_\alpha \approx 5 \times 10^8$ for a $1\, 000$ particle simulation
at $\phi = 0.6$. This value 
agrees well with the BW fit that gives $4.85 \times 10^8$ for $\phi=0.6$, but is orders of magnitude different
than the prediction of the VFT fit, $1.1 \times 10^{10}$ for $\phi = 0.6$. While we expect a $1\, 000$ particle
system to be too small for $\phi=0.6$ and, in particular, we expect that at this volume fraction 
$\xi(\tau_\alpha)$ is larger than half of the $1\, 000$ particle system simulation cell,
this result does provide some evidence for the BW fit.  
However, large, fully equilibrated simulations at
$\phi$ larger than those utilized in our simulations are needed to test the proper functional form of the divergence. 

We should emphasize here that our simulations cover both the mode-coupling-like regime and a new
regime in which the data are consistent with $\ln(\tau_\alpha) \sim (\phi_0 - \phi)^{-2}$. 

\section{Calculation of $\chi_4(t)$ and $\xi(t)$}\label{xicalc}
Here we describe the details of the calculation of the dynamic susceptibility $\chi_4(t)$ and 
the dynamic correlation length $\xi(t)$. The outline of this appendix is as follows.
First, we present a derivation of Eq.~\eqref{chicorrection}. Next, we show that $\chi_4(t)$ calculated from this equation
agrees with an independent extrapolation of $S_4(q;t)$ to $q=0$. Finally, we use 
$\chi_4(t)$ calculated from formula ~\eqref{chicorrection} 
as the point at $S_4(q=0;t)$ in a fitting procedure to find the most reliable result for 
the dynamic susceptibility and the dynamic correlation length.

In our simulation, the volume fraction and the concentration of particles is fixed. Thus, fluctuations of the
volume fraction and concentration do not contribute to the direct calculation of the dynamic susceptibility.
To account for these fluctuations we follow the procedure
introduced by Lebowitz \textit{et al.} \cite{Lebowitz1967}. We start by considering an ensemble where the number of
particles can fluctuate and calculate the first order corrections to $\chi_4(t)$ calculated in
an ensemble where the volume fraction and the concentration are held constant. Consider
\begin{eqnarray}\label{chimu}
\lefteqn{\chi_4(t) = } &&\nonumber \\&&
\frac{1}{\left< N \right>_{(\mu_1,\mu_2,V)}} 
\left< \delta \left( \sum_n w_n(t)\right) \delta \left( \sum_m w_m(t) \right) \right>_{(\mu_1,\mu_2,V)}, \nonumber \\
\end{eqnarray}
where $\left< \right>_{\mathbf{X}}$ denotes an ensemble where $\mathbf{X}$ is held constant. In
Eq.~\ref{chimu} the chemical potentials 
of both small and large particles, $\mu_1$ and $\mu_2$, and the volume ($V$) are held constant, 
but the numbers of particles, $N_1$ and $N_2$, are allowed to fluctuate. Since the volume
is held constant in all the ensembles considered, we will not explicitly indicate the constant $V$ in what follows. 
According to Eq.~2.11 of Lebowitz \textit{et al.}\ \cite{Lebowitz1967},
\begin{eqnarray}\label{chifull}
\left<N\right>_{\mu_1,\mu_2} \chi_4(t) 
& = & \left< \delta \left(\sum_n w_n(t) \right) \delta \left( \sum_m w_m(t) \right) \right>_{N_1,N_2} 
\nonumber \\
&& + \left< (\delta N_1)^2 \right>_{\mu_1,\mu_2} 
\left[ \frac{\partial \left< \sum_n w_n \right>_{N_1,N_2}}{\partial N_1} \right]^2 \nonumber \\
&& + \left< (\delta N_2)^2 \right>_{\mu_1,\mu_2} 
\left[ \frac{\partial \left< \sum_n w_n \right>_{N_1,N_2}}{\partial N_2} \right]^2 \nonumber \\
&& + 2 \left< \delta N_1 \delta N_2 \right>_{\mu_1,\mu_2} \nonumber \\
&& \times \frac{\partial \left< \sum_n w_n \right>_{N_1,N_2}}{\partial N_1} 
\frac{\partial \left< \sum_n w_n \right>_{N_1,N_2}}{\partial N_2}
.
\end{eqnarray}
We utilize the relationship 
\begin{eqnarray}
\left<N\right>_{\mu_1,\mu_2}^{-1} \left< \delta N_n \delta N_m \right>_{\mu_1,\mu_2} 
&=& \lim_{q \rightarrow 0} \sqrt{x_n x_m} S_{nm}(q) \nonumber \\ &=& \sqrt{x_n x_m} S_{nm},
\end{eqnarray} 
where $S_{nm}(q)$ is the partial structure factor and $x_n = N_n/N$. 
We also recognize that $\left< \sum_n w_n(t) \right>_N = (N_1 + N_2) F_o(t)$.
Finally, we replace the differentiation with respect to the numbers of particles with the differentiation with respect
to the volume fraction and the concentration and in this way we obtain
\begin{eqnarray}
\chi_4(t) & = & \chi_4(t)_{N} + \chi_\phi^2 H_1 + \chi_\phi \chi_c H_2 + \chi_c^2 H_3 \nonumber \\
&& + F_o(t)^2 H_4 + F_o(t) \chi_\phi H_5 + F_o(t) \chi_c H_6,
\nonumber \\
\end{eqnarray}
where $\chi_x = \partial F_o(t)/\partial x$. The $H_n$ are functions of $S_{nm}$, and are given by
\begin{eqnarray}
H_1 & = & \left( \frac{\pi \rho}{6} \right)^2 \left[ d_1^6 x_1 S_{11} + 2 d_1^3 d_2^3 \sqrt{x_1 x_2} S_{12} + d_2^6 x_2 S_{22} 
\right] 
\nonumber \\ \label{H1}
\\
H_2  & = & \frac{\pi \rho}{3} \left[d_1^3 x_1 x_2 S_{11} - d_1^3 x_1 \sqrt{x_1 x_2} S_{12} \right.
\nonumber \\
&& \left. + d_2^3 x_2 \sqrt{x_1 x_2} S_{12} - d_2^3 x_1 x_2 S_{22} \right] 
\\
H_3 & = & x_2^2 x_1 S_{11} - 2 x_1 x_2 \sqrt{x_1 x_2} S_{12} + x_1^2 x_2 S_{22}
\\
H_4 &= &x_1 S_{11} + 2 \sqrt{x_1 x_2} S_{12} + x_2 S_{22}
\\
H_5 &=& \frac{\pi \rho}{3} \left[ d_1^3 x_1 S_{11} + (d_1^3 + d_2^3) \sqrt{x_1 x_2} S_{12} + d_2^3 x_2 S_{22} \right]
\nonumber \\ 
\\
H_6 &=& 2\left[ x_1 x_2 S_{11} + (x_2-x_1) \sqrt{x_1 x_2} S_{12} - x_1 x_2 S_{22} \right].
\nonumber \\
\label{H6}
\end{eqnarray}

To calculate $H_n$ we fit the wave vector dependent version of $H_n$, \textit{i.e.} expressions \eqref{H1}-\eqref{H6}
with $S_{nm}$ replaced by $S_{nm}(q)$, to a wave-vector independent constants for $q \le 0.6$. 
Due to noise in our data we cannot perform a more accurate extrapolation. We
checked this approach by using the same procedure to calculate the pressure using
the partial structure factors. We checked that the pressure obtained from the $q\to 0$ limit of the 
structure factors agrees with the pressure obtained from the extrapolation of the pair correlation function to contact. 

\begin{figure}
\includegraphics[width=3.2in]{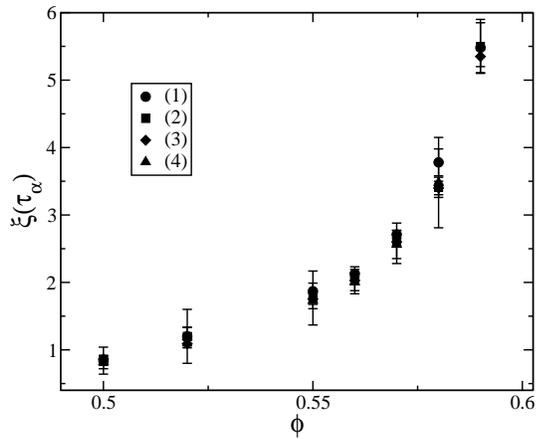}
\caption{\label{xifit}The dynamic correlation length $\xi(\tau_\alpha)$ 
obtained using the different fitting procedures described
in the text. The number correspond to the different fitting procedures.}
\end{figure}

\begin{figure}
\includegraphics[width=3.2in]{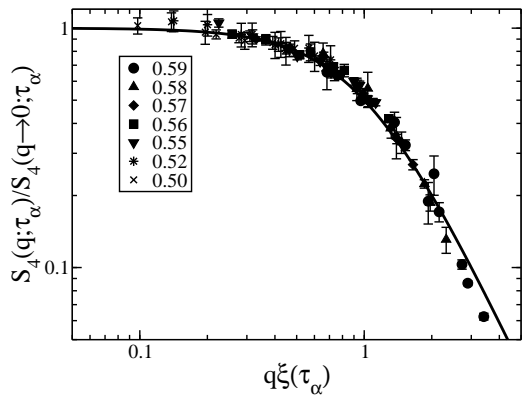}
\caption{\label{s4scale}Scaling plot $S_4(q;\tau_\alpha)/S_4(q\to 0;\tau_\alpha)$ versus $q \xi(\tau_\alpha)$ 
for the $80\, 000$ particle simulations.}
\end{figure}

To verify Eq.~\eqref{chifull} and to check its accuracy, we extrapolated $S_4(q;t)$ to $q \rightarrow 0$
by fitting $S_4(q;t)$ obtained for non-zero wave-vectors to an Ornstein-Zernicke function. We 
compared the resulting $\lim_{q\to 0} S_4(q;t)$ to $\chi_4(t)$ obtained from Eq.~\eqref{chifull}. The extrapolation agreed to within
error and thus we concluded that Eq.~\eqref{chifull} provided a good means to calculate $\chi_4(t)$. Subsequently, we 
used Eq.~\eqref{chifull} as the $q=0$ value of $S_4(q;t)$ in fitting procedures. 

It is important to recognize that the above described verification of Eq.~\eqref{chifull} requires a rather large system size.
In particular, we could only perform it using $N=80\, 000$ particles system. Once using $\chi_4(t)$ obtained from Eq.~\eqref{chifull}
as the $q=0$ value of $S_4(q;t)$ is accepted, we were able to use moderately large systems ($N=10\, 000$ particles).

We fit $S_4(q;t)$, using Eq.~\eqref{chifull} as the $q=0$ value of $S_4(q;t)$, to several functions of the form
\begin{equation}
S_4(q;t) = \frac{A}{1+(\xi q)^2 + C^2 q^4} + \frac{B}{(1 + (\xi q)^2)^2},
\end{equation}
where all the fitting parameters are time dependent.
We performed the following fits: (1) set $C=0$ and $B=0$, i.e.\ an Ornstein-Zernicke type fit;
(2) set B=0 which gives a function suggested by the inhomogeneous mode-coupling theory \cite{Biroli2006}; 
(3) set $A=\chi_4(t)|_{\phi,c}$ and $C=0$, which results in a function suggested by
field theoretic considerations \cite{Berthier2007p2,Berthier2007p3}. We also fit $S_4(q;t)$ 
to a function utilized by Stein and Andersen
\cite{Stein2008}, $\ln[S_4(q;t)] = \ln(A) - [\xi q]^2 + C q^4$, procedure (4). All of the fits results except for procedure (3) results
in statistically the same length, Fig.~\ref{xifit}, if we restrict the fits as follows. For procedure (1), the 
Ornstein-Zernicke fits, we only fit to $q \le 1.5/\xi$ and for the fit to $\ln[S_4(q;t)]$ we only fit 
$q \le 1/\xi$. Procedure (3) resulted in an $\xi$ approximately 1.2 times smaller than the other procedures at
every volume fraction, thus none of the conclusions of this work changes due to utilizing that fitting function.
For volume fractions beyond our ability to study, it may be found that $\xi$ determined through procedures
(1), (2), and (4) is not simply a factor of $\xi$ found using procedure (3). As a final check, we used $\xi$ obtained
from the Ornstein-Zernicke fit to check the quality of overlap of $S_4(q;\tau_\alpha)/S_4(q \to 0;\tau_\alpha)$
versus $q \xi(\tau_\alpha)$, Fig.~\ref{s4scale}, and we find the overlap to be very good. 
The results shown in Figs.~\eqref{s4}, \eqref{chixi}, \eqref{xitime}, \eqref{xi2time}, \eqref{tauDxi},
and \eqref{s4scale} are found by the Ornstein-Zernicke fits. 


\end{document}